\documentclass[12pt, draftclsnofoot, onecolumn]{IEEEtran}

\usepackage{ifpdf}

  \usepackage[dvips]{graphicx}
  \graphicspath{{../eps/}}
  \DeclareGraphicsExtensions{.eps}

\usepackage{mathbbol}
\usepackage{cite}
\usepackage{algorithm}
\usepackage{algorithmic}
\usepackage[cmex10]{amsmath}
\usepackage{amsthm}
\usepackage{amssymb}
\usepackage{balance}
\usepackage{eqparbox}
\usepackage{color}
\usepackage[cmintegrals]{newtxmath}

\newtheorem{theorem}{Theorem}
\newtheorem{remark}{Remark}

\newtheorem{proposition}{Proposition}
\newcommand{\bs}[1]{\boldsymbol{#1}}

\newcommand{\mb}[1]{\mathbf{#1}}
\newcommand{\mr}[1]{\mathrm{#1}}

\newcommand{\bseq}{\begin{subequations}}
\newcommand{\eseq}{\end{subequations}}
\newcommand{\baln}{\begin{align}}
\newcommand{\ealn}{\end{align}}
\newcommand{\balnd}{\begin{aligned}}
\newcommand{\ealnd}{\end{aligned}}
\newcommand{\beq}{\begin{equation}}
\newcommand{\eeq}{\end{equation}}
\newcommand{\beqn}{\begin{eqnarray}}
\newcommand{\eeqn}{\end{eqnarray}}
\newcommand{\beqno}{\begin{eqnarray*}}
\newcommand{\eeqno}{\end{eqnarray*}}
\newcommand{\bma}{\begin{displaymath}}
\newcommand{\ema}{\end{displaymath}}
\newcommand{\bnu}{\begin{enumerate}}
\newcommand{\enu}{\end{enumerate}}
\newcommand{\bce}{\begin{center}}
\newcommand{\ece}{\end{center}}
\newcommand{\btb}{\begin{tabular}}
\newcommand{\etb}{\end{tabular}}
\newcommand{\ba}{\begin{array}}
\newcommand{\ea}{\end{array}}
\title{System Energy-Efficient Hybrid Beamforming for mmWave Multi-user Systems}
\author{Vu Nguyen Ha, {\em Member, IEEE,} Duy H. N. Nguyen, {\em Member, IEEE,} \\ and Jean-Fran\c{c}ois Frigon, {\em Senior Member, IEEE}
\thanks{Vu N. Ha and Jean-Fran\c{c}ois Frigon are with \'{E}cole Polytechnique de Montr\'{e}al, Poly-Grames Research Center, Montreal, Quebec, Canada, H3T 1J4 (e-mail: \{vu.ha-nguyen,j-f.frigon\}@polymtl.ca).}
\thanks{Duy H. N. Nguyen is with Department of Electrical and Computer Engineering, San Diego State University, San Diego, CA, USA 92182 (e-mail: duy.nguyen@sdsu.edu).}
}

\begin{document}
\maketitle

\begin{abstract}
This paper develops energy-efficient hybrid beamforming designs for mmWave multi-user systems where analog precoding is realized by switches and phase shifters such that radio frequency (RF) chain to transmit antenna connections can be switched off for energy saving. By explicitly considering the effect of each connection on the required power for baseband and RF signal processing, we describe the total power consumption in a sparsity form of the analog precoding matrix. However, these sparsity terms and sparsity-modulus constraints of the analog precoding make the system energy-efficiency maximization problem non-convex and challenging to solve. To tackle this problem, we first transform it into a subtractive-form weighted sum rate and power problem. A compressed sensing-based re-weighted quadratic-form relaxation method is employed to deal with the sparsity parts and the sparsity-modulus constraints. We then exploit alternating minimization of the mean-squared error to solve the equivalent problem where the digital precoding vectors and the analog precoding matrix are updated sequentially. The energy efficiency upper bound and a heuristic algorithm are also examined for comparison purposes. Numerical results confirm the superior performances of the proposed algorithm over benchmark energy-efficiency hybrid precoding algorithms and heuristic one.
\end{abstract}


\section{Introduction}
\IEEEPARstart{R}{\lowercase{e}cently}, mmWave has been considered as a promising technology for emerging wireless networks to deal with  the increasing wireless traffic demands \cite{Rappaport_2013,PiKhan2011,Andrews_JSAC_2014}.
Operating in the frequency bands from 30-300 GHz, this technology can empower multi-Gbps transmission speed.   
Thanks to the band's short wavelength, a large number of antenna elements can be leveraged in a small space at the transceivers. Hence, multiple data streams for multiple users can be transmitted via spatial multiplexing which potentially results in a significant improvement in spectral efficiency \cite{Roh14,Ayach2014}.

Employing one single RF chain for each antenna as in the conventional fully digital precoder design typically requires high implementation cost and complexity \cite{Werthmann13}. Thus, hybrid precoding (HP) has been proposed as a cost-efficient beamforming technique for the mmWave system \cite{Yu_TSP_16,Alkhateeb-COMMag-2014,Alkhateeb-TWC2014,Alkhateeb-TCom16,Bogale_TWC16,Duy-Access17}. This proposed transceiver architecture adds the analog precoder (AP) to the conventional digital precoder (DP); hence, the number of RF chains can be reduced to save the operation cost \cite{Ayach2014,Alkhateeb-TWC2014,Bogale_TWC16,Duy-Access17,Duy-ICC16,Duy-TWC17,Duy-TWC18}.
Additional components typically consists of analog phase shifters connecting RF chains to antennas to achieve analog beamforming gain. Interestingly, HP enables near-optimal performance thanks to the low-rank characteristics of mmWave channels \cite{Bogale_TWC16}. 
However, the design of mmWave transceivers still raises concerns on the system energy efficiency (SEE), which is an important aspect of mmWave systems \cite{Prasad_17,Yazdan17,Du18}.
Besides the conventional achievable rate and transmit power trade-off, the study in HP implementation for mmWave system should cover the impact of the design structure on the SEE.


There are two main HP structures, named fully-connected and sub-connected \cite{Du18}. The first structure activates all the phase shifters between RF chains and antennas while only a subset of phase shifters is activated for analog signal processing in the second structure.
Many studies on these two structures have been reported that turning on larger set of phase shifters can achieve the high capacity by enhancing more degree-of-freedom; however, employing the large number of phase shifters in AP component may result in high prohibitive power consumption. In addition, turning off a number of phase shifters can also lessen the hardware complexity by lowering the number of RF paths since each RF chain is allowed to connect to a subset of all antennas, which hence can ease the implementation and drop down the consumed energy \cite{Ahmed_CST18}. Therefore, the design of efficient HP taking care of optimizing the subset of phase shifters for SEE maximization in mmWave multiple-input and multiple-output (MIMO) systems is an interesting and challenging problem, which is the focus of this paper. 

\subsection{Related Works}
While research on HP for mmWave systems is plentiful, limited work has studied maximizing the SEE. Several papers have considered the energy-efficient HP designs for mmWave systems or massive MIMO systems without optimizing the SEE, such as \cite{Gao_jsac_16,Zhang_Tcom_17}. 
In particular, an energy-efficient HP design for mmWave MIMO systems is proposed in \cite{Gao_jsac_16} based on the successive interference cancellation method. A hybrid analog-digital architecture for HP in mmWave MIMO systems is proposed in \cite{Zhang_Tcom_17}, where multiple sub-arrays are employed at the transmit and receive antennas.
Both works investigate SEE achieved by the proposed HP designs. 
HP design maximizing the SEE is directly studied in \cite{Zi_Jsac_16,He_Tsp_17,Gao_ICC_17}.
Specifically, \cite{Zi_Jsac_16} develops a novel HP design to maximize the SEE of massive MIMO system.
In this work, the SEE is formulated as the ratio between the achievable rate and the total power consumption which is the sum of transmit power and constant components. The upper-bound fully digital precoding (FDP) is first optimized, based on which the HP is reconstructed by minimizing the Euclidean distance between two precoding designs. The numbers of RF chains and antennas are then optimized based on the statistical analysis values of SEE for very large array antenna systems.
Optimizing the number of RF chains is also considered in \cite{He_Tsp_17} to maximize the SEE of a HP mmWave system. 
In this work, the power radiation is formulated as a linear function of the number of RF chains, then, an efficient codebook-based hybrid precoding design is proposed by jointly selecting the AP in a codebook set and optimizing the baseband ones.
Using a different method, Gao \emph{et al.} \cite{Gao_ICC_17} studied the SEE HP for mmWave massive MIMO system by employing machine learning tools. In particular, a hardware-efficient analog network structure has been developed in this work where a new group-connected mapping strategy for HP is introduced. Among predetermined activated phase-shifter groups corresponding to different hardware implementations, the most efficient group together with the corresponding HP is then selected.  

To the best of our knowledge, the SEE HP design for mmWave multi-user system which optimizes all the numbers of utilized RF chains, transmission antennas, and the connections among the RF chains and the antennas has been not studied in the literature.
Filling this gap, this paper studies a novel HP structure for mmWave MIMO system where each of the connections between RF chains and antennas can be optimally activated or deactivated by utilizing ON-OFF switches to maximize the SEE.

\subsection{Research Contributions}
We consider a fully-connected HP structure where a switch is integrated with a  phase shifter over every RF chain to antenna connection. The switches can be activated or deactivated to save the power consumption of the phase shifters. In addition, a RF chain (or/and an antenna) can be deactivated for energy saving if all of its corresponding connections are turned off. This hardware structure is transferred into sparsity-modulus constraints for AP matrix design. In particular, the absolute value of an AP matrix element can be one or zero. The sparsity-modulus design also enables us to exploit the total power consumption as a sparsity function of AP matrix. The SEE is then calculated as the ratio between the achievable rate and total consumption power based on which the SEE maximization (SEEM) problem is formulated as a non-convex problem. In preliminary work reported in \cite{VuHa_ICC18}, we proposed an iterative algorithm to tackle this optimization problem without a detailed proof of convergence. Performance comparison between the proposed algorithm and heuristic algorithms was not also given. In solving the SEEM problem and exposing energy-efficient HP designs, the contributions of our paper are given as follows:
\begin{itemize}
\item To tackle the SEEM problem, we first transform it into a subtractive-form weighted sum rate and power (WSRP) problem based on the \textit{``Dinkelbach's method''} \cite{Dinkelbach67}. Then, we exploit an alternating minimization of the mean-squared error (MMSE) algorithm to solve the WSRP problem where the DP vectors and AP matrix are updated alternatively. In each iteration, we employ compressed sensing-based relaxation method to deal with the sparsity-modulus constraints and sparse formulation of total power consumption. In particular, this method help transform the MMSE problem into a convex one, for which a locally optimal solution can be found efficiently. The analysis on the convergence of the proposed algorithm is also given.
\item The energy efficiency upper bound and an heuristic algorithm are also studied for comparison purpose. Extensive numerical studies are conducted where we examine the convergence and efficiency of the proposed algorithms as well as the impacts of different system parameters on the SEE.
\end{itemize} 

The remaining of this paper is organized as follows. We describe the system model, and formulations of the SEEM HP design problem in Section~\ref{sec:SM}. 
In Section~\ref{sec2}, we transfer the SEEM problem into the WSRP problem and point out the general design algorithm.
The compress-sensing-based method is then proposed in Sections~\ref{sec3} to solve the WSRP problem based on which we developed the novel HP design algorithm to maximize the SEE. 
The upper bound of SEE and an heuristic algorithm are presented in Section~\ref{sec4}. 
Numerical results are presented in Section~\ref{sec5} followed by conclusions in Section~\ref{ccls}. 

\emph{Notations:}  $(\mathbf{X})^{T}$ and $(\mathbf{X})^{H}$ denote the transpose and conjugate transpose of the matrix $\mathbf{X}$, respectively; $\Vert\mb{x}\Vert_0$ and $\Vert\mb{x}\Vert$ denote the norm-$0$ and Euclidean norm of a vector $\mb{x}$, respectively.

\section{System Model} \label{sec:SM}
\subsection{Multi-user Hybrid Precoding System Model} \label{ssec:MUHBSM}
\begin{figure*}[!t]
	\centering
	\includegraphics[width=165mm]{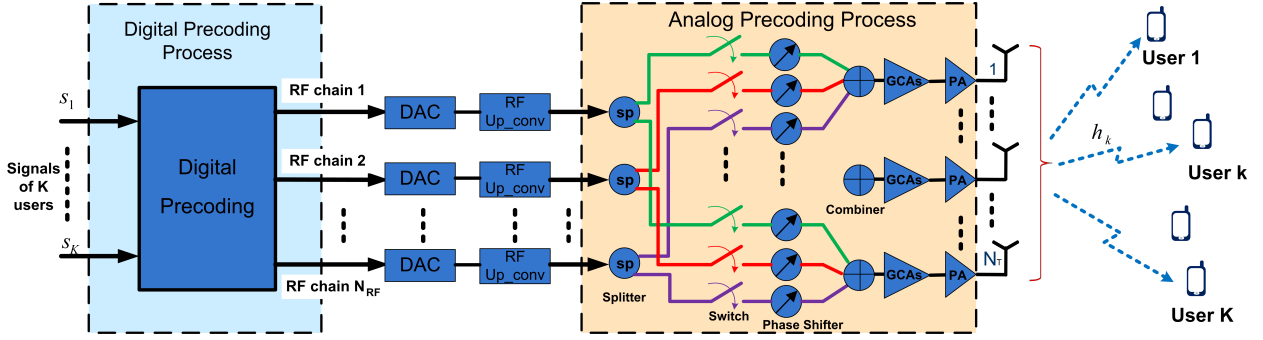}
	\caption{Diagram of a mmWave multi-user system with hybrid analog/digital precoding design.}
	\label{hybrid-fig}
\end{figure*}

Consider a downlink mmWave multi-user HP system where a base station (BS) equipped with $N_{\mr{T}}$ antennas and $N_{\mr{RF}}$ RF chains serves $K$ remote single-antenna users. Utilizing the HP, the BS first applies the DP vectors to the corresponding symbol sequences for the users. Specifically, a DP vector $\mb{w}_{k} \in \mathbb{C}^{N_{\mr{RF}} \times 1}$ is applied to the data symbol $s_{k} \in \mathbb{C}$, intended for user $k$. Without loss of generality, we assume $|s_{k}| = 1$. Following the digitally precoded sequences, the BS then employs an AP matrix, $\mb{A} \in \mathbb{C}^{N_{\mr{T}} \times N_{\mr{RF}}}$ , to map the RF signals from $N_{\mr{RF}}$ RF chains to $N_{\mr{T}}$ antennas. 
{\color{blue}Let $a_t^n$ be the element allocated on the $t^{th}$ row and the $n^{th}$ column of $\mb{A}$ and $x_n$ be the signal on RF chain $n$ which can be determined as
$x_n = \sum_{\forall k} w_k^n s_k$ where $w_k^n$ is the $n^{th}$ element of DP vector $\mb{w}_{k}$.
Then, the signal of RF chain $n$ at antenna $t$ after going through the AP block can be written as $x_n a_t^n$.
In this work, we consider a dynamic fully-connected RF chains to antennas structure in which $\mb{A}$ is implemented by integrating switches and the phase shifters. The ON-OFF switch deployed on each RF chain to antenna connection can allow (or disallow) the corresponding RF signal be forwarded (or not be forwarded) to that antenna for transmission.
When the connection between RF chain $n$ and antenna $t$ is  deactivated (turned off), we can set $\big|a_t^n\big|^2 = 0$.
Inversely, this connection is activated (turned on), the corresponding RF signal will be phase shifted, combined to others, and transmitted by antenna $t$. In this case, $\big|a_t^n\big|^2$ should be $1$, and the phase-shifted version of $x_n$ will be $x_n e^{j \theta_t^n}$ where $a_t^n = e^{j \theta_t^n}$.
Hence, the following sparsity-modulus condition for the elements of $\mb{A}$ can help the AP matrix mathematically present well the implementation of switches and phase shifters in our system.
\beq
\big|a_t^n\big|^2 = 1 \text{ or } 0 \quad \forall (t,n).
\eeq}
By taking into account of the HP design for multi-user system in \cite{Alkhateeb-TWC2014}, the signal received by user $k$ can be given as
\begin{equation} \label{recv_sig}
y_k =  \mb{h}_{k}^H \sum_{\forall j} \mb{A} \mb{w}_j s_j + n^{\mr{noise}}_k,
\end{equation} 
where $n^{\mr{noise}}_k$ is the additive Gaussian noise at user $k$ and $\mb{h}_{k} \in \mathbb{C}^{N_{\mr{T}}}$ is the multiple-input and single-output (MISO) channel from the BS to user $k$. 
Assuming coherent detection at the users, the signal-to-interference-plus-noise ratio (SINR) at user $k$ can be given as
\begin{eqnarray}\label{SINR}
\mr{SINR}_{k} = \frac{ \big|\mb{h}^H_{k} \mb{A} \mb{w}_{k} \big|^2}{\sum_{j \neq k} \big|\mb{h}^H_{k} \mb{A} \mb{w}_{j} \big|^2 + \sigma^2},
\end{eqnarray}
where $\sigma^2$ is the power of additive Gaussian noise. 
Assuming Gaussian signaling between the BS and the users, the total achievable data-rate of the system can be described as
\beq \label{rate}
R(\mb{W},\mb{A}) = \sum_{\forall k} \log(1 + \mr{SINR}_{k}),
\eeq
where $\mathbf{W} = [\mathbf{w}_1,\ldots,\mathbf{w}_K]$ is denoted as the matrix generated by all DP vectors. 
{\color{blue} \begin{remark}
		The ON-OFF switches employed in the mmWave transmitter allow a selection of termination (inactive port) or pass-through (active port) for the RF chain signals \cite{R_Cory09}.
	Each switch is matched to an output of one splitter which is implemented to divide a RF chain signal to $N_{\sf{T}}$ antennas as shown in Fig.~\ref{hybrid-fig}.
	Once the active port is selected, the signal is passed through, which is indicated as ON state.  
	On another hand, when an inactive port is selected by the switch, the OFF state occurs.
	In the considered system, a good match is assumed, which means that a matched termination is implemented at the inactive port and the inactive signal is terminated by 50-Ohm load \cite{R_Cory09}.
\end{remark}}

{\color{blue}\subsection{Power Consumption Model} \label{ssec:ECM}
	In this section, the power consumption model is analyzed by counting the required power of each system component. In general, the total power consumption in the system is comprised of the power consumed by the  digital signal processing (DSP) hardware, the RF signal processing hardware, and the RF signal radiation \cite{Hien_book16}.
	\subsubsection{DSP Power Consumption}
	For the DSP component, the static power consumption corresponding to each user's signal is due to parts of the baseband signal process. 
	In this paper, we assume that the power consumption for the DSP hardware is unchanged, which is given by
	\beq
	P_{\mr{DP}} = K P_{\mr{BB}},
	\eeq
	where $P_{\mr{BB}}$ represents the power consumption for the baseband signal processing of one user.
	\subsubsection{Power Consumption by RF Signal Processing Hardware}
	As illustrated in Fig.~\ref{hybrid-fig}, the baseband signals are first converted to an analog signals and up-converted to RF band. 
	Then, an $(N_{\sf{T}}+1)$-port splitter (so called divider) is implemented over each RF chain to divide the RF signal to $N_{\sf{T}}$ outputs corresponding to $N_{\sf{T}}$ antennas. 
	Each of these $N_{\sf{T}}$ outputs will be matched to an ON-OFF switch.
	Here, the signal will be terminated or passed to the phase-shifter accordingly if $\vert a_t^n \vert^2 = 0 \text{ or } 1$, respectively. 
	All signals heading to one antenna are then combined by employing an $(N_{\sf{RF}}+1)$-port combiner. The combined signal is passed through the gain-compensation amplifier (GCA) and the power amplifier (PA) before being propagated by that very antenna. 	Denote $P_{\mr{DAC}}$, $P_{\mr{RFC}}$, $P_{\mr{SW}}$ and $P_{\mr{PS}}$ as the power consumption of the digital-to-analog converter (ADC), RF converter, ON-OFF switch, and phase shifter, respectively. 
	In practice, the power consumed by each of these components can be assumed to be unchanged \cite{Rial15}.
	However, the power consumed by the amplifiers varies due to the loss caused by the splitters as well as combiners \cite{V_Jamali_ICC19}, and due to the transmission power.
	\paragraph{Power Consumption by GCAs}
	In this system, a number of GCAs are deployed to boost the RF signals, which lost part of their power after passing through splitters, switches, phase shifters and combiners; and to attain sufficient power for driving the PAs at the antennas. 
	To estimate the GCA's power consumption, we revisit the power loss of each connection between RF chains and antennas which is caused by splitters and combiners.
	Assume that a multi-port splitter (or combiner) is implemented based on a cascade of simple three-port splitters (or combiners) as in \cite{V_Jamali_ICC19}. Then, the losses of the outputs of $(N_{\sf{T}}+1)$-port splitter and $(N_{\sf{RF}}+1)$-port combiner can be estimated as  $\lceil \log_2(N_{\sf{T}}) \rceil L_{\sf{d}}$ and $\lceil \log_2(N_{\sf{RF}}) \rceil L_{\sf{c}}$ \cite{V_Jamali_ICC19} where $\lceil . \rceil$ stands for ceiling function and $L_{\sf{d}}$ and $L_{\sf{c}}$ (in dB) represent the power losses of a three-port splitter and a three-port combiner, respectively.
	In addition, denote $L_{\sf{sw}}$ and $L_{\sf{ps}}$ (in dB) as the losses when RF signal passes through the switch and phase shifter. 
	Let $G_{\sf{amp}}$ (in dB) be the maximum amplification gain of one GCA. Then, the number of GCAs required for compensating signal before going to a PA can be calculated as
	\beq
	M_{\mr{GCA}} = \left\lceil \dfrac{\lceil \log_2(N_{\sf{T}}) \rceil L_{\sf{d}} + \lceil \log_2(N_{\sf{RF}}) \rceil L_{\sf{c}} + L_{\sf{sw}} + L_{\sf{ps}}}{G_{\sf{amp}}}\right\rceil,
	\eeq  
	Then, the power consumption of the GCAs corresponding to one activated antenna is given by
	\beq
	P_{\mr{GCA}} = M_{\mr{GCA}} P_{\mr{amp}},
	\eeq
	where $P_{\mr{amp}}$ represents the power consumption of one GCA.
	\paragraph{Power Consumption by PAs}
	The power consumption of PAs can be modeled in a linear form of the radiation power \cite{V_Jamali_ICC19,H_Yan_CSM19}.
	In particular, the power consumption by the PA implemented at antenna $t$, named $P_{\mr{PA},t}$, can be calculated as
	\beq
	P_{\mr{PA},t} = P_t/\rho_{\mr{pa}},
	\eeq
	where $P_t$ represents the transmission power at antenna $t$ and $\rho_{\mr{pa}}$ stands for the power amplifier efficiency \cite{H_Yan_CSM19}.
	
	\paragraph{Power Consumption of RF Signal Processing Hardware}
	An RF-chain-to-antenna connection is activated or deactivated by the ON-OFF switch allocated on the corresponding connecting link. An activated connection would consume certain amount of power for the RF chain and the antenna connected to it, as illustrated in Fig.~\ref{hybrid-fig}. Furthermore, one RF chain can be turned off for power saving if there is no connection from that RF chain to any antenna. Likewise, an antenna may be inactive when all connections from RF chains to that antenna are turned off. It is recalled that $|a_t^n|^2 = 1$ implies an active connection between RF chain $n$  and antenna $t$ and $|a_t^n|^2 = 0$ implies otherwise.
	Hence, we can express the total power consumption for RF signal processing in a sparse form of $\mb{A}$ as follows:
	\beq \label{RF_power_r12}
	P_{\mr{RF}} = (P_{\mr{DAC}}+P_{\mr{RFC}}) \Vert \mb{c}(\mb{A})\Vert_0 + N_{\mr{T}}N_{\mr{RF}}P_{\mr{SW}} + P_{\mr{PS}} \Vert \mb{A} \Vert_0 + P_{\mr{GCA}} \Vert \mb{r}(\mb{A})\Vert_0 + \frac{1}{\rho_{\mr{pa}}}\sum_{\forall t} P_{t},
	\eeq
	where $\mb{c}(\mb{A}) = [c_1,\ldots,c_{N_\mr{RF}}]^T \in \mathbb{C}^{N_{\mr{RF}}}$ and $c_n =\sum_{\forall t} |a_t^n|^2$,
	$\mb{r}(\mb{A}) = [r_1,...,r_{N_\mr{T}}]^T \in \mathbb{C}^{N_{\mr{T}}}$ and $r_t =\sum_{\forall n} |a_t^n|^2$. 
	
	\subsubsection{Transmission Power and Total Power Consumption}
	Next, the transmission power can be calculated based on $(\mb{W},\mb{A})$ as follows:
	\beq
	P_{\mr{T}}(\mb{W},\mb{A})= \sum_{\forall k} \mb{w}^H_{k}\mb{A}^H\mb{A}\mb{w}_{k}.
	\eeq
	Then, total power consumption can be calculated by taking the summation of $P_{\mr{DP}}$, $P_{\mr{RF}} (\mb{A})$, and $P_{\mr{T}}$.
	In addition, the last component in \eqref{RF_power_r12} represents the total transmission power over all antennas, which yields $\sum_{\forall t} P_t = \sum_{\forall k} \mb{w}^H_{k}\mb{A}^H\mb{A}\mb{w}_{k}$. Hence, the system power consumption can be described as
	\beqn
	P_{\mr{tot}}(\mb{W},\mb{A}) & = & P_{\mr{DP}}+ N_{\mr{T}}N_{\mr{RF}}P_{\mr{SW}}  + (P_{\mr{DAC}}+P_{\mr{RFC}}) \Vert \mb{c}(\mb{A})\Vert_0 + P_{\mr{PS}} \Vert \mb{A} \Vert_0 + P_{\mr{GCA}} \Vert \mb{r}(\mb{A})\Vert_0 \nonumber \\
	& & \hspace{8cm} + \left( 1 + \frac{1}{\rho_{\mr{pa}}} \right) \sum_{\forall k} \mb{w}^H_{k}\mb{A}^H\mb{A}\mb{w}_{k} \nonumber \\
	& = & P_{\mr{cons}} + P_{\mr{spr}}(\mb{A}) + \left( 1 + \frac{1}{\rho_{\mr{pa}}} \right) P_{\mr{T}}(\mb{W},\mb{A}).
	\eeqn
	where $P_{\mr{cons}} \! = \! P_{\mr{DP}} \! + \! N_{\mr{T}}N_{\mr{RF}}P_{\mr{SW}}$ and 
	$ P_{\mr{spr}}(\mb{A}) \! = \! (P_{\mr{DAC}} \! + \! P_{\mr{RFC}}) \Vert \mb{c}(\mb{A})\Vert_0 \! + \! P_{\mr{PS}} \Vert \mb{A} \Vert_0 \! + \! P_{\mr{GCA}} \Vert \mb{r}(\mb{A})\Vert_0$.}

\subsection{Problem Formulation}
We now ready to define the SEE (in bits/Hz/W) as the ratio of the achievable sum-rate to the total power consumption as follows.
\beq
\eta(\mb{W},\mb{A})=\dfrac{R(\mb{W},\mb{A})}{P_{\mr{tot}}(\mb{W},\mb{A})}.
\eeq
\begin{figure*}[t!]
	\centering
	\includegraphics[width=160mm]{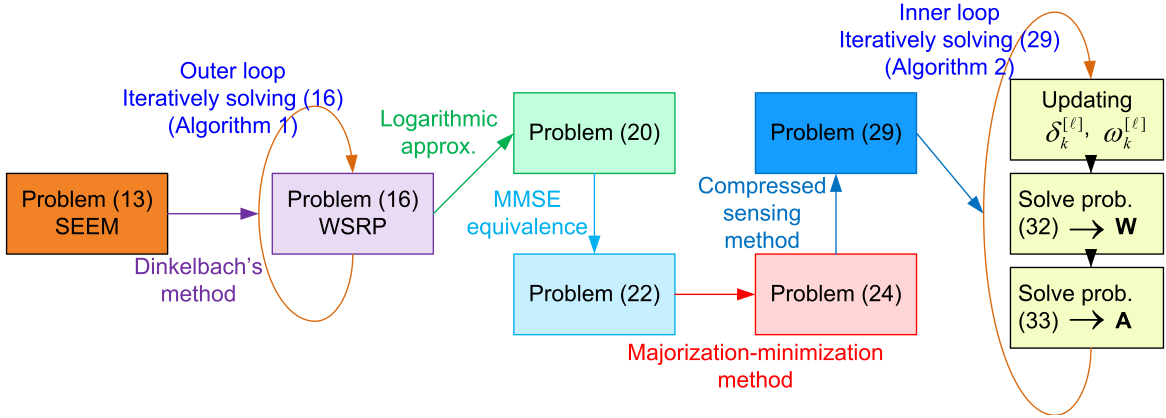}
	\caption{Diagram of solution approach.}
	\label{flow-chart-fig}
\end{figure*}
In this paper, we are interested in jointly optimizing the DP vectors and the sparsity-modulus AP matrix to maximize the SEE under the constraint on the transmit power budget at each antenna. This SEEM problem can be stated as
\begin{subequations} \label{SEE-max-prb}
\begin{eqnarray} 
\hspace{-0.3cm}&\underset{\mb{W},\mb{A}}{\max}& \eta(\mb{W},\mb{A})=\dfrac{\sum_{\forall k} \log(1 + \mr{SINR}_{k})}{P_{\mr{cons}} + P_{\mr{spr}}(\mb{A}) + \left( 1 + \frac{1}{\rho_{\mr{pa}}} \right) P_{\mr{T}}(\mb{W},\mb{A})} \\
\hspace{-0.3cm}&\text{s. t. }& \big|a_t^n\big| = 1 \text{ or } 0, \forall (t,n), \label{cnt1} \\
\hspace{-0.3cm}&& \sum_{\forall k} \mb{w}^H_{k}\mb{a}_t \mb{a}_t^H\mb{w}_{k} \leq P^{\mr{max}}_t, 1 \leq t \leq N_{\mr{T}}, \label{cnt2}
\end{eqnarray}
\end{subequations}
where $\mb{a}_{t} \in \mathbb{C}^{N_{\mr{RF}} \times 1}$ is denoted as the vector corresponding to the $t^{th}$ column of $\mb{A}^H$, and $P^{\mr{max}}_t$ is transmit power limit at antenna $t$ of the BS.
{\color{blue}The challenges for solving problem (\ref{SEE-max-prb}) come from the 
fractional form of the objective function and the sparsity terms associated with the power consumption model.
To overcome these challenges, we first apply Dinkelbach's method \cite{Dinkelbach67} for solving problem (\ref{SEE-max-prb}) by transforming it into a sequence of parameterized subtractive-form problems.
Then, compressed sensing based solution approaches \cite{Candes08} are employed to dual with the sparsity terms in the parameterized problems. In addition, MMSE-based transformation \cite{Christensen08} and majorization-minimization based method \cite{Hunter04} are also enhanced for proposing a framework solving problem (\ref{SEE-max-prb}). The overall solution approach is summarized in Fig.~\ref{flow-chart-fig}, in which each of stages will be given in the subsequent sections.}

\section{General Design Algorithm based on Dinkelbach's Method}
\label{sec2}
\subsection{Transformation of SEE Maximization Problem}
{\color{blue} The objective of problem \eqref{SEE-max-prb} is in a fractional form, which is difficult to tackle directly. 
Dinkelbach in \cite{Dinkelbach67} has proposed an efficient method for solving optimization problem with this type of objective function, which is well-known as the Dinkelbach algorithm.
Specifically, the method first transforms the fractional problem into a parameterized subtractive form. An iterative solution approach to update the parameter of subtractive-form problem is then invoked to obtain its optimal solution. 
The foundation of this method can be given in the following theorem which summarizes the theoretical results in \cite{Dinkelbach67}. 

\begin{theorem} \label{thr_eta_star}
Consider two following problems
\beqn
(\mathcal{P}_I) & \underset{\mb{x}}{\max} & R(\mb{x})/P(\mb{x}) \quad \text{s. t. } \mb{x} \in \mathcal{S}, \\
(\mathcal{P}_{II}^{\eta}) & \underset{\mb{x}}{\max} & R(\mb{x}) - \eta P(\mb{x}) \quad \text{s. t. } \mb{x} \in \mathcal{S}. \label{sub-x-prob}
\eeqn
which can represent any arbitrary fractional and subtractive forms, respectively.
Let $\chi(\eta)$ be the optimal objective value of \eqref{sub-x-prob} for given $\eta$, which can be considered as a function of $\eta$. Assume that $P(\mb{x}) > 0$ $\forall \mb{x} \in \mathcal{S}$ and $\eta^{*}$ is the optimal objective value of $(\mathcal{P}_I)$. Then, there are four observations as follows.
\begin{itemize}
\item[i)] $\chi(\eta)$ is a strictly monotonic decreasing function.
\item[ii)] $\chi(\eta) > 0$ if and only if $\eta < \eta^{*}$.
\item[iii)] $\chi(\eta) < 0$ if and only if $\eta > \eta^{*}$.
\item[iv)] $(\mathcal{P}_I)$ and $(\mathcal{P}_{II}^{\eta^{*}})$ have the same set of optimal solutions. 
\end{itemize}
\end{theorem}
\begin{IEEEproof}
The proof is given in Appendix~\ref{prf_thr_eta_star}.
\end{IEEEproof} 

Theorem~\ref{thr_eta_star} has provided the foundation, based on which the transformation of SEEM problem can be performed
and an algorithmic solution approach can be devised to solve the SEEM problem. First, let us consider the weighted power (WSRP) maximization problem
for given value of $\eta$ which is stated as follows.
\beq
\underset{(\mb{W},\mb{A})}{\max}\bar{\chi}(\eta,\mb{W},\mb{A}) = R(\mb{W},\mb{A})- \eta P_{\mr{tot}}(\mb{W},\mb{A}) \text{ s. t. } (\mb{W},\mb{A}) \in \Theta,
\label{sub-tract-prob}
\eeq
where $\Theta$ stand for the feasible set of $(\mb{W},\mb{A})$ in SEEM problem. 
As can be seen, parameter $\eta$ in the WSRP problem acts as a negative weight
on the total energy consumption; hence,  it can be considered
viewed as the ``price'' of the system's energy consumption which can be adjusted to meet the efficient point of our design. 
We also denote $\eta^{\star}$ as the maximum SEE which can be expressed as
\beqn \label{eta-star}
\eta^{\star} =  \eta(\mb{W}^{\star},\mb{A}^{\star})   =  \dfrac{R(\mb{W}^{\star} ,\mb{A}^{\star})}{P_{\mr{tot}}(\mb{W}^{\star},\mb{A}^{\star})} = \underset{(\mb{W},\mb{A}) \in \Theta}{\max} \dfrac{R(\mb{W},\mb{A})}{P_{\mr{tot}}(\mb{W},\mb{A})},
\eeqn
where $(\mb{W}^{\star},\mb{A}^{\star})$ represents the optimal solution. Then, the observation (iv) in Theorem~\ref{thr_eta_star} yields that SEEM problem and WSRP with $\eta^{\star}$ have the same set of optimal solutions. Hence, the SEEM problem can be solved by iteratively addressing the solution of WSRP problem for a certain value of $\eta$ and adjusting $\eta$ until an optimal $\eta^{\star} \geq 0$ satisfying $\chi(\eta^{\star})=0$ is found. }

\subsection{Overview of the Solution Approach}
In this section, we present an overview of the proposed solution approach which is summarized in Algorithm ~\ref{P2_alg:1}. The algorithm relies on updating $\eta$ and solving the corresponding WSRP problem iteratively based on the well-known Dinkelbach-type solution method \cite{Dinkelbach67,Crouzeix85}.  We start by setting $\eta^{(0)}=0$. 
In iteration $m$, corresponding to a certain value $\eta^{(m)}$, the WSRP problem is solved to achieve the optimal value $(\mb{W}^{(m)},\mb{A}^{(m)})$.
Then, the value of $\eta$ is updated for the next iteration as $\eta^{(m+1)}=\frac{R(\mb{W}^{(m)},\mb{A}^{(m)})}{P_{\mr{tot}}(\mb{W}^{(m)},\mb{A}^{(m)})}$. According to the proof given in \cite{Dinkelbach67,Crouzeix85}, this process ensures the monotonic increase of $\eta^{(m)}$  if the WSRP problem is solved optimally in each iteration, which yields the convergence of Algorithm~\ref{P2_alg:1}. However, obtaining such an optimal solution is challenging due to the non-convexity of WSRP problem. 
The following theorem establishes the convergence of Algorithm~\ref{P2_alg:1} when only local optimal solutions are found at each iteration.

\begin{algorithm}[!t]
\caption{\textsc{Overview of the Proposed Algorithm}}
\label{P2_alg:1}
\begin{algorithmic}[1]
\STATE Initialize $\eta^{(0)}=0$, set $m=0$, and choose predetermined tolerate $\tau^{\mr{out}}$.
\REPEAT 
\STATE Solve \eqref{sub-tract-prob} with $\eta^{(m)}$ to achieve $(\mb{W}^{(m)},\mb{A}^{(m)})$.
\STATE Set $\eta^{(m+1)}=\frac{R(\mb{W}^{(m)},\mb{A}^{(m)})}{P_{\mr{tot}}(\mb{W}^{(m)},\mb{A}^{(m)})}$.
\STATE Update $m:=m+1$.
\UNTIL $\vert \eta^{(m)} - \eta^{(m-1)} \vert \leq \tau^{\mr{out}}$.
\STATE Return $(\mb{W}^{(m-1)},\mb{A}^{(m-1)})$.
\end{algorithmic}
\end{algorithm}

\begin{theorem} \label{P2:prp_cvgAlg1}
For a given $\eta^{(m)}$ in the $m$-th iteration of Algorithm~\ref{P2_alg:1}, if a local optimal solution of WSRP problem, $(\mb{W}^{(m)},\mb{A}^{(m)})$, is found such that 
\beq \label{cvg_cdt_Alg1}
\bar{\chi}(\eta^{(m)},\mb{W}^{(m)},\mb{A}^{(m)}) \geq \bar{\chi}(\eta^{(m)},\mb{W}^{(m-1)},\mb{A}^{(m-1)}),
\eeq
Algorithm~\ref{P2_alg:1} will converge after a finite number of iterations.
\end{theorem}
\begin{IEEEproof}
The proof is given in Appendix~\ref{prf_P2_prp_cvgAlg1}. 
\end{IEEEproof}

The next task is to find an efficient method to solve the WSRP problem. This problem is NP-hard as a result of the
non-convex sum rate, the $\ell_0$-norm of matrix $\mb{A}$ in the objective function, and the sparsity-modulus constraint \eqref{cnt1}. Hence, finding its globally optimal solution is prohibitively complex. 
Theorem~\ref{P2:prp_cvgAlg1} encourages us to develop an efficient (probably sub-optimal) solution to indicate a local optimal solution of the WSRP problem in each iteration of Algorithm~\ref{P2_alg:1}.

{\color{blue}
\begin{remark}
	It is worth noting that the bisection searching approaching can be applied for updating $\eta^{(m)}$ in each iteration of Algorithm~\ref{P2_alg:1} instead of utilizing \textbf{Step 4} since $\chi(\eta)$ is monotonic decreasing function respect to $\eta$. However, bisection searching method, while being simple in the context, is quite complex the implementation,  since there is no exact method of finding the initial upper and lower bounds.
	In addition, existing studies in literature \cite{Z_Dai_ECCV12,C_Wu_WCL16} have demonstrated that updating $\eta^{(m)}$ as in \textbf{Step 4} based on Dinkelbach method can converge faster than bisection approaches in many circumstances. 
\end{remark}}

\section{Energy-Efficiency Hybrid Precoding Design}
\label{sec3}
{\color{blue}This section presents an efficient solution approach to the WSRP problem via  CS methods. In particular, the sparsity terms in $P_{\mr{spr}}$ are first transformed into the approximated continuous logarithmic forms. Then, the CS method is employed to relax the sparsity-modulus constraints of variables $a_t^n$'s, elements of $\mb{A}$. 
To do so, the WSRP problem containing sparsity can be relaxed to non-sparsity problem which can be solved efficiently by iteratively dealing with number of quadratically constrained quadratic programming problems.
The details are given as follows.}

\subsection{Sparsity Relaxation and MMSE-based Transformation} \label{sec:spar_rlx}
It is worth to note that the $\ell_0$-norms of matrix $\mb{A}$, $\mb{r}(\mb{A})$, and $\mb{c}(\mb{A})$ can be defined directly from $\Vert a_t^n \Vert_0$. Interestingly, the sparsity solutions can be obtained by employing the re-weighted $\ell_1$-norm minimization methods, originally proposed to enhance the data acquisition in compressed sensing. In particular, for a given non-negative real vector $\mb{v} \in \mathbb{R}_{+}^{M}$, its $\ell_0$-norm can be approximated as 
\beq
\Vert \mb{v} \Vert_0 = \underset{\varepsilon \rightarrow  0}{\lim}\sum \dfrac{\log(1+|v_i|\varepsilon^{-1})}{\log(1+\varepsilon^{-1})},
\eeq
where $\varepsilon \ll 1$. 
\newcounter{tempequationcounter}
\begin{figure*}[!t]
\normalsize
\setcounter{tempequationcounter}{\value{equation}}
\begin{IEEEeqnarray}{rCl}
&\underset{ \mb{A},\mb{W}}{\max} &\;
  l(\mb{W},\mb{A}) = \kappa \sum_{\forall k} \log(1 + \mr{SINR}_{k}) - \kappa \eta (1 + \dfrac{1}{\rho_{\mr{pa}}})\sum_{\forall k} \mb{w}_{k}^H \mb{A}^H \mb{A} \mb{w}_{k}  \nonumber \\
&& - \eta\bigg[ \kappa P_{\mr{cons}} + P_{\mr{PS}}\sum \limits_{\forall (t,n)} \log(|a_t^n|^2+\varepsilon) + P_{\mr{GCA}} \sum \limits_{\forall n} \log(c_n + \varepsilon)  + (P_{\mr{DAC}}+P_{\mr{RFC}})\sum \limits_{\forall t} \log(r_t + \varepsilon )\bigg]  \nonumber \\
&\mathrm{s.t.} &\; \textrm{ constraints \eqref{cnt1} and \eqref{cnt2}.} \label{WSRP_appx} 
\end{IEEEeqnarray}
\setcounter{equation}{\value{tempequationcounter}}
\hrulefill
\vspace*{4pt}
\end{figure*}
Then, the WSRP problem can be approximated to problem \eqref{WSRP_appx}\addtocounter{equation}{1} at the top of page \pageref{WSRP_appx} where $\kappa=\log(1+\varepsilon^{-1})$. To simplify the objective function \eqref{WSRP_appx}, we define $P_{\mr{Log}}(\mb{W},\mb{A})$ as 
\beq \label{l0_appx}
P_{\mr{Log}}(\mb{W},\mb{A}) = \sum_{\forall X}P_X\log(X+\varepsilon),
\eeq
where $X$ stands for $\{|a_t^{n}|^2\}$'s, $\{c_n\}$'s, and $\{r_t\}$'s, and $P_X$ represents to $P_{\mr{PS}}$, $P_{\mr{GCA}}$, and $(P_{\mr{DAC}}+P_{\mr{RFC}})$.
Then, we can address the non-convex problem (\ref{WSRP_appx}) by relating it to a weighted sum-mean square error (MSE) minimization problem as mentioned in the following theorem.

{\color{blue}\begin{theorem}
\label{P2_thr2}
Problem \eqref{WSRP_appx} is equivalent to the following
weighted sum-MSE minimization problem, i.e., two problems have same optimal solutions,
\begin{eqnarray}
\hspace{-0.3cm} &\underset{ \mb{A},\mb{W},\left\lbrace\delta_{k},\omega_{k}\right\rbrace}{\min} &
g(\mb{W},\mb{A},\delta_{k},\omega_{k}) = \kappa  \sum \limits_{\forall k} \left(\omega_{k} e_k -  \log \omega_{k}  -  1 \right) \nonumber \\
&& \hspace{4.5cm} +  \eta \bigg[\kappa  (1 + \dfrac{1}{\rho_{\mr{pa}}}) \sum_{\forall k} \mb{w}_{k}^H \mb{A}^H \mb{A} \mb{w}_{k} + \kappa P_{\mr{cons}}  +  P_{\mr{Log}}(\mb{W},\mb{A})\bigg]  \nonumber \\
&\mathrm{s.t.} & \textrm{ constraints \eqref{cnt1} and \eqref{cnt2}.} \label{WMMSE-prob2}
\end{eqnarray}
where $e_k = \mathbb{E} \left[ \big|s_{k} - \delta_{k} y_{k}\big|^2 \right]$, $\omega_{k}$ and $\delta_{k}$ represent the MSE weight and the receive coefficient for user $k$, respectively.
\end{theorem}}
\begin{IEEEproof} 
The proof is given in Appendix~\ref{prf_P2_thr2}. 
\end{IEEEproof}

\subsection{MMSE-based Hybrid Precoding Design}
\subsubsection{Majorization-Minimization based Method}
{\color{blue} As can be observed, $P_{\mr{Log}}(\mb{W},\mb{A})$ term in objective function of problem \eqref{WMMSE-prob2} is the summation of concave functions which makes this minimization problem hard to be tackled.
To overcome this challenge, one can employ the well-known majorization-minimization solution approach which implements an iteration algorithm to solve the optimization problem \cite{Hunter04,Y_Sun_TSP17,M_Figueiredo_TIP07,Najafi_GLC17}. 
The principle of this algorithm consists of the following processes in each iteration: (i) approximating the objective function to tackle-able forms at a predetermined fixed point; (ii) solving the approximate problems; (iii) updating the fixed point for the next iteration based on the recent achieved optimal solution.
Employing this approach to solve problem \eqref{WMMSE-prob2}, let $X^{(\ell)}$ represent a fixed value of $X$ of the $\ell$-th iteration.
Then, the concavity of $\log(X+\varepsilon)$ function can be majored by the following result at $X^{(\ell)}$:
\beq \label{log_maj}
\log(X+\varepsilon) \leq \log\big(X^{(\ell)}+\varepsilon\big)+\frac{1}{X^{(\ell)}+\varepsilon}\big(X - X^{(\ell)}\big),
\eeq 
This result implies that for given value of $X^{(\ell)}$, the right hand side (RHS) of \eqref{log_maj} can be employed as a majoring function for $\log(X+\varepsilon)$.
Then, in the $\ell$-th iteration, the majorization-minimization method \cite{Hunter04,Y_Sun_TSP17} focuses on minimizing the approximate problem of \eqref{WMMSE-prob2} which is formed by replacing $\log(X+\varepsilon)$ terms by the RHS of \eqref{log_maj} \cite{Hunter04,Y_Sun_TSP17}. Specifically, an iterative solution can be employed where problem \eqref{WMMSE-prob2} can be approximated to the following problem by replacing \eqref{log_maj} into \eqref{WMMSE-prob2} at iteration $(\ell+1)$.
\begin{eqnarray} 
\hspace{-0.3cm}&\underset{ \mb{A},\mb{W},\left\lbrace\delta_{k},\omega_{k}\right\rbrace}{\min}& \tilde{g}(\mb{W},\mb{A},\delta_{k}, \omega_{k}) = \kappa \sum \limits_{\forall k} \left(\omega_{k} e_k - \log \omega_{k} - 1 \right) \nonumber \\
&& \hspace{3.5cm} +\eta\bigg[\kappa P_{\mr{cons}} + \sum \limits_{t=1}^{N_{\mr{T}}} \mb{a}_t^H \mb{D}_t^{[\ell]}  \mb{a}_t  + \kappa (1 + \dfrac{1}{\rho_{\mr{pa}}}) \sum \limits_{t = 1}^{N_{\mr{T}}} \mb{a}_t^H  \big(  \sum_{\forall k} \mb{w}_{k} \mb{w}_{k}^H  \big) \mb{a}_t\bigg]   \nonumber \\
\hspace{-0.3cm}&\text{s. t. }& \text{ constraints \eqref{cnt1} and \eqref{cnt2}}. \label{WMMSE-prob}
\end{eqnarray}
In \eqref{WMMSE-prob}, $\mb{D}_t$ is calculated based on the outcomes of the previous iteration as
\beq \label{Dt}
\mb{D}_t^{[\ell]}=P_{\mr{PS}}\Psi_t^{[\ell]} + P_{\mr{GCA}} \varphi_t^{[\ell]} \mb{I} + (P_{\mr{DAC}}  +  P_{\mr{RFC}})\mb{F}_t\sum_{\forall n} \phi_n^{[\ell]},
\eeq
where $\Psi_t^{[\ell]}=diag(\psi_t^{1,[\ell]},...,\psi_t^{N_{\mr{RF}},[\ell]})$, $\mb{F}_t=diag(0_{1 \times (t-1)},1$ $,0_{1 \times (N_{\mr{RF}}-t)})$, and  $\mb{I}$ is the identify matrix with the size of $N_{\mr{RF}} \times N_{\mr{RF}}$. In \eqref{Dt}, $\psi_t^{[\ell]}$, $\phi_n^{[\ell]}$, and $\varphi_t^{[\ell]}$ are the weighting factors corresponding to $|a_t^{n,[\ell]}|^2$, $c_n^{[\ell]}$, and $r_t^{[\ell]}$ achieved from $\ell$-th iteration, which are updated based on \eqref{log_maj} as
\beqn  \label{weight_factor}
\psi_t^{n,[\ell]} = \dfrac{1}{|a_t^{n,[\ell]}|^2 + \varepsilon}, \quad  \phi_n^{[\ell]}  =  \dfrac{1}{c_n^{[\ell]} + \varepsilon}, \quad \varphi_t^{[\ell]}  = \dfrac{1}{r_t^{[\ell]} + \varepsilon}. 
\eeqn
By properly choosing and updating $\psi_t$'s, $\phi_n$'s, and $\varphi_t$'s iteratively, the non-convex discontinuous $\ell_0$-norms can be effectively approximated to the quadratic forms \cite{Candes08}.
This has confirmed that replacing the sparsity term $\Vert \mb{x} \Vert_0$ by $\sum_{\forall i} \left( {1}/{\vert x_i \vert^2 +\varepsilon} \right)^{1/2} x_i$ is useful to develop an algorithm
solving sparsity problem which is also a very efficient benchmark approach in CS employed in many works in literature \cite{Candes08,Tony_Quek_TWC13,WeiYu_Access14}.

Problem \eqref{WMMSE-prob} is however still very challenging due to the discontinuous constraint \eqref{cnt1}. 
As can be seen, this challenging issue can be transferred to the following 
\beq
\vert a_t^n \vert = \Vert a_t^n \Vert_0 \quad \forall (t,n).
\eeq
which is also in a sparsity form.
Regarding CS technique, we can relax constraint \eqref{cnt1} in the iteration $(\ell+1)$ by approximating $a_t^n$ as
\beq
a_t^n = \left( {1}/{|a_t^{n,[\ell]}|^2 + \varepsilon} \right)^{1/2} a_t^n = (\psi_t^{n,[\ell]})^{1/2} a_t^n.
\eeq 
Then, problem \eqref{WMMSE-prob} can be further approximated as
\begin{subequations} \label{WMMSE-probII}
\begin{eqnarray} 
\hspace{-0.3cm}&\underset{ \mb{A},\mb{W},\left\lbrace\delta_{k},\omega_{k}\right\rbrace}{\min}& \tilde{g}(\mb{W},\mb{A},\delta_{k}, \omega_{k}) = \kappa \sum \limits_{\forall k} \left(\omega_{k} e_k - \log \omega_{k} - 1 \right)  \nonumber \\
&& \hspace{.3cm} +\eta\left[\kappa P_{\mr{cons}}  + \sum \limits_{t=1}^{N_{\mr{T}}} \mb{a}_t^H \mb{D}_t^{[\ell]} \Psi_t^{[\ell]} \mb{a}_t  + \kappa (1 + \dfrac{1}{\rho_{\mr{pa}}}) \sum \limits_{t = 1}^{N_{\mr{T}}} \mb{a}_t^H \Psi_t^{[\ell]} \bigg(  \sum_{\forall k} \mb{w}_{k} \mb{w}_{k}^H  \bigg) \mb{a}_t\right]   \\
\hspace{-0.3cm}&\text{s. t. }& \vert a_t^n \vert \leq 1 \quad \forall (t,n), \label{cnt1b} \\
&& \mb{w}^H_{k} \Psi_t^{[\ell]} \mb{a}_t \mb{a}_t^H  \mb{w}_{k} \leq P^{\mr{max}}_t, 1 \leq t \leq N_{\mr{T}}. \label{cnt2b} 
\end{eqnarray}
\end{subequations}

It is noted that the objective function in problem \eqref{WMMSE-probII} is not \emph{jointly} convex, but it is convex over each set of variables $\mb{w}_{k}$'s, $\mb{a}_{t}$'s, $\delta_{k}$'s, and $\omega_{k}$'s. Hence, an efficient algorithm for solving this problem can be developed by alternately optimizing $\mb{w}_{k}$'s and $\mb{a}_{t}$'s, and the MSE weight update for $\delta_{k}$'s, and $\omega_{k}$'s. 

\subsubsection{Update MSE Weights and Receive Coefficients}
For given $(\mb{W},\mb{A})$, $\delta_{k}$'s, and $\omega_{k}$'s can be determined according to the results in Appendix~\ref{prf_P2_thr2}. In particular, the MMSE receive filter at user $k$ is given as
\begin{eqnarray} \label{receive2}
\delta_{k}^{\star} = \delta_{k}^{\mr{MMSE}} = \bigg(\sum_{\forall j} \vert\mb{h}_{k}^{H} \mb{A} \mb{w}_j\vert^2 +  \sigma_k^2 \bigg)^{-1}\mb{w}_k^H \mb{A}^H \mb{h}_{k}. 
\end{eqnarray}
And, the optimum value of $\omega_k$ can be expressed as
\beqn \label{omega2}
\omega_k^{\star} \! = \! e_k^{-1} \! = \! 1 \! + \! \bigg(\sum_{\forall j} \vert\mb{h}_{k}^{H} \mb{A} \mb{w}_j\vert^2 +  \sigma_k^2 \bigg)^{-1}\!\!\!\!\vert \mb{w}_k^H \mb{A}^H \mb{h}_{k}\vert^2.
\eeqn
\subsubsection{Digital Precoding Design}
For given AP matrix $\mb{A}$, the optimal $\mb{w}_{k}$'s can be obtained by solving the following QCQP:
\begin{eqnarray}
\hspace{-0.3cm} & \underset{\left\lbrace \mb{w}_{k}\right\rbrace}{\min} &
\sum \limits_{\forall (k)}  \mb{w}_{k}^H \Big[\sum_{\forall j} \omega_{j} |\delta_{j}|^2 \mb{A}^H \mb{h}_{j} \mb{h}_{j}^H \mb{A} + \eta (1 + \dfrac{1}{\rho_{\mr{pa}}}) \sum_{\forall t} \Psi_t^{[\ell]} \mb{a}_t \mb{a}_t^H  \Big] \mb{w}_{k} \nonumber \\
& & \hspace{6cm} - \omega_{j} \delta_k^{\prime}\mb{w}_k^{H} \mb{A}^H \mb{h}_k - \omega_{j} \delta_k \mb{h}_k^{H} \mb{A} \mb{w}_k    \nonumber \\
\hspace{-0.3cm} & \mathrm{s.t.} & \textrm{\; constraint \eqref{cnt2b}.} \label{QCQP-probW}
\end{eqnarray}
This QCQP problem can be solved by any standard convex optimization solvers or the  Lagrangian duality method.
\subsubsection{Sparsity-Modulus Analog Phase-Shifting Matrix Design}
For given DP vectors $\mb{w}_k$'s, the AP matrix $\mb{A}$ can be achieved by solving the following problem:
\begin{eqnarray} \label{QCQP-probA}
\hspace{-0.3cm} & \underset{\left\lbrace \mb{a}_{t}\right\rbrace}{\min}&
 \sum \limits_{t = 1}^{N_{\mr{T}}}  \Big(   \mb{a}_{t}^H  \Lambda_t \mb{a}_{t}    + \eta \mb{a}_t^H  \Omega_{\mb{w}}  \mb{a}_t + \eta\mb{a}_t^H  \mb{D}_t^{[\ell]} \Psi_t^{[\ell]}  \mb{a}_t  -  \sum_{\forall k}\omega_{k} \delta_k^{\prime} h_{k,t} \mb{w}_k^{H} \mb{a}_t \! - \! \mb{a}_t^H \! \sum_{\forall k} \! \omega_{k} \delta_k h_{k,t}^{\prime} \! \mb{w}_k \! \Big) \nonumber \\
\hspace{-0.3cm} & \mathrm{s.t.} & \textrm{\; constraints  \eqref{cnt1b} and \eqref{cnt2b},} 
\end{eqnarray}
where $\Lambda_t^{[\ell]} = \sum_{\forall k} \omega_{k} |\delta_{k}|^2 |h_{k,t}|^2 \Omega_{\mb{w}}^{[\ell]}$ and $\Omega_{\mb{w}}^{[\ell]}=\kappa(1 + \dfrac{1}{\rho_{\mr{pa}}})\Psi_t^{[\ell]} \sum_{\forall k} \mb{w}_{k} \mb{w}_{k}^H$. 
 As can be observed, problem \eqref{QCQP-probA} can be decomposed into $N_{\mr{T}}$ simpler sub-problems corresponding to $N_{\mr{T}}$ vectors $\mb{a}_t$'s. Specifically, the sub-problem corresponding to $\mb{a}_t$ can be stated as follows:
\begin{subequations} \label{QCQP-probAt}
\begin{eqnarray} 
(\mathcal{P}_t) & \underset{ \mb{a}_{t}}{\min}&
\mb{a}_{t}^H  \Phi_t \mb{a}_{t}  -  \mb{f}_t^H \mb{a}_t  -  \mb{a}_t^H \mb{f}_t  \\
& \mathrm{s.t.} & |a_t^n| \leq 1, \;\; \forall n,\\
&& \mb{a}_{t}^H \Psi_t^{1/2} \Omega_{\mb{w}} \Psi_t^{1/2} \mb{a}_{t} \leq P_{t}^{\mr{max}},
\end{eqnarray}
\end{subequations}
where $\Phi_t^{[\ell]} = \Lambda_t^{[\ell]} + \eta\Omega_{\mb{w}}^{[\ell]}+\eta\mb{D}_t^{[\ell]}\Psi_t^{[\ell]} $ and $\mb{f}_t = \sum_{\forall k}\omega_{k} \delta_k h_{k,t}^{\prime} \mb{w}_k^{H}$. Again, problem $(\mathcal{P}_t)$ is a QCQP-form problem which can be solved by employing any standard optimization solver. }

\subsubsection{MMSE-based Hybrid Precoding Design}
By iteratively updating $\left\lbrace \mb{w}_{k}, \mb{a}_{t}, \delta_{k}, \omega_k \right\rbrace$, we can obtain the
MMSE hybrid precoding. Combined with the compressed sensing-based method and with updating the weight factors $\psi_{t}$'s, $\varphi_t$'s, and $\phi_n$'s, the DP vectors and sparsity-modulus AP matrix design for WSRP maximization is summarized in Algorithm~\ref{P2_alg:2}. The convergence of this algorithm is analyzed in the following theorem.
\begin{algorithm}[!t]
\caption{\textsc{Iterative Weighted Rate and Power Maximization Hybrid Precoding}}
\label{P2_alg:2}
\begin{algorithmic}[1]
\STATE Initialize: Choose suitable $\mb{w}_{k}^{[0]}$'s and $\mb{A}^{[0]}$, set $\ell=0$ and a predetermined tolerate $\tau^{\mr{in}}$.
\REPEAT 
\STATE Update $\psi_t^{[\ell]}$'s, $\phi_n^{[\ell]}$'s, $\varphi_t^{[\ell]}$'s as in \eqref{weight_factor}.
\STATE Calculate $\delta_{k}^{[\ell+1]}$'s as in \eqref{receive2}.
\STATE Calculate and $\omega_{k}^{[\ell+1]}$'s as in and \eqref{omega2}.
\STATE Determine $\mb{w}_{k}^{[\ell+1]}$'s by solving problem (\ref{QCQP-probW}) corresponding to $\mb{A}^{[\ell]}$'s, $\delta_{k}^{[\ell]}$'s and $\omega_{k}^{[\ell]}$'s.
\FOR{$t=1$ to $N_{\mr{T}}$}
\STATE Solve problem (\ref{QCQP-probAt}) corresponding to $\mb{W}^{[\ell+1]}$'s, $\delta_{k}^{[\ell]}$'s and $\omega_{k}^{[\ell]}$'s to determine $\mb{a}_{t}^{[\ell+1]}$.
\ENDFOR
\STATE Update $\ell:=\ell+1$.
\UNTIL $\vert \tilde{g}(\mb{W}^{[\ell]},\mb{A}^{[\ell]},\bs{\delta}^{[\ell]},\bs{\omega}^{[\ell]}) - \tilde{g}(\mb{W}^{[\ell -1]},\mb{A}^{[\ell -1]},\bs{\delta}^{[\ell -1]},\bs{\omega}^{[\ell -1]})\vert \leq \tau^{\mr{in}}$.
\STATE Return $(\mb{W},\mb{A})$ as $\mb{W}^{[\ell]}$ and $a_t^{n[\ell]}=a_t^{n[\ell]} \sqrt{\psi_t^{n[\ell]}}$, $\forall(t,n)$.
\end{algorithmic}
\end{algorithm}

\begin{theorem} \label{prp_cvgAlg2}
Algorithm~\ref{P2_alg:2} converges to a local optimum solution after a finite number of iterations.
\end{theorem}
\begin{IEEEproof} 
The proof is given in Appendix~\ref{prf_P2_prp_cvgAlg2}.
\end{IEEEproof}

\subsection{Energy-Efficiency Hybrid Precoding Implementation with Algorithms~\ref{P2_alg:1} and \ref{P2_alg:2}}
Algorithm~\ref{P2_alg:2} can be employed in Step 3 of Algorithm~\ref{P2_alg:1} to solve the SEEM problem.
To ease the notation, we denote the repeated steps of Algorithm~\ref{P2_alg:1} as the outer loop, and that due to Algorithm~\ref{P2_alg:2} as the inner loop.
In this implementation, we choose the outcome of the $(m-1)$-th outer iteration, $(\mb{W}^{(m-1)},\mb{A}^{(m-1)})$, as the initial point for running Algorithm~\ref{P2_alg:2} in the next outer iteration. 
This careful selection ensures the convergence condition of Algorithm~\ref{P2_alg:1} in Theorem~\ref{P2:prp_cvgAlg1}, which is analyzed in the following proposition.
\begin{proposition} \label{P2_prp_cvgAlg1_Alg2}
Consider the implementation where Algorithm~\ref{P2_alg:2} is employed in the Step 3 of Algorithm~\ref{P2_alg:1}. 
If in every outer iteration, such as the $m$-th one, initial point of of the inner loop running Algorithm~\ref{P2_alg:2} is set as $(\mb{W}^{(m-1)},\mb{A}^{(m-1)})$, then, the condition \eqref{cvg_cdt_Alg1} is satisfied.
\end{proposition}
\begin{IEEEproof} 
The proof is given in Appendix~\ref{prf_P2_prp_cvgAlg1_Alg2}.
\end{IEEEproof}

\section{Upper Bound of System Energy Efficiency and Heuristic Solution}
\label{sec4}
{\color{blue}This section focuses on developing a framework which can exploit the existing HP designs to solve problem \eqref{SEE-max-prb}.
Specifically, a two-stage heuristic algorithm is proposed based on works given in \cite{Zi_Jsac_16} and \cite{Ayach2014}.
First, the upper bound of SEE is analyzed according to modifying an algorithm presented in \cite{Zi_Jsac_16} designing the FDP to maximize the ratio between the achievable rate and transmission energy. The FDP obtained in this stage is so-called upper-bound FDP.
In the next stage, an heuristic iterative solution is developed by step-by-step updating the RF chain to antenna connection structure.
Particularly, in each iteration, one link between RF chains and antennas is considered to be turned off if after deactivating it, the updated connection structure together with a near-optimal performance HP re-constructed from the upper-bound FDP based on the method given in \cite{Ayach2014} can improve the SEE. 

\subsection{Upper-Bound of System Energy Efficiency}
\label{sec:upper}
This section introduces the upper bound of SEE and a framework obtaining the upper-bound FDP.
Let $\mb{B} \in \mathbb{R}^{N_{\mr{T}} \times N_{\mr{RF}}}$ be the mapping matrix, where the elements of $\mb{B}$ represent whether the RF chain to antenna connections are activated or not. 
Specifically, the mapping matrix corresponding to given AP matrix $\mb{A}$ can be defined as
\beq \label{a_b_related}
b_t^n = | a_t^n |^2, \forall (t,n),
\eeq
where $b_t^n$ be the element allocated on the $t^{th}$ row and the $n^{th}$ column of $\mb{B}$. According to constraint \eqref{cnt1}, we have that $b_t^n$ can be $1$ or $0$. 
In addition, the total power consumption according to the RF process can be expressed based on $\mb{B}$ as
\beqn
P_{\mr{spr}}(\mb{A})&=&\bar{P}_{\mr{spr}}(\mb{B})= (P_{\mr{DAC}}+P_{\mr{RFC}}) \Vert \mb{c}(\mb{B})\Vert_0 + P_{\mr{PS}} \Vert \mb{B} \Vert_0 + P_{\mr{GCA}} \Vert \mb{r}(\mb{B})\Vert_0.
\eeqn
Hence, for given matrix $\mb{B}$, if \eqref{a_b_related} holds, the SEE can be rewritten as
\beq
\eta(\mb{W},\mb{A})=\dfrac{\sum_{\forall k} \log(1 + \mr{SINR}_{k})}{P_{\mr{cons}}+\bar{P}_{\mr{spr}}(\mb{B})+(1+1/\rho_{\mr{pa}})\sum_{\forall k} \mb{w}_k^H\mb{A}^H\mb{A}\mb{w}_k}.
\eeq
Then, the upper bound of $\eta(\mb{W},\mb{A})$ can be determined based on the following proposition.
\begin{proposition} \label{P2_prp1}
	When $\mb{A}$ satisfies \eqref{a_b_related}, $\eta(\mb{W},\mb{A})$ can be upper bounded by
	\beq
	\bar{\eta}(\mb{U}^\mr{Up},\mb{B}) = \dfrac{\sum_{\forall k} R^{\mr{FDP}}_{k}(\mb{U}^{\mr{Up}})}{P_{\mr{cons}}+\bar{P}_{\mr{spr}}(\mb{B})+ (1+1/\rho_{\mr{pa}}) \sum_{\forall k} \mb{u}_k^{\mr{Up} H}\mb{u}^{\mr{Up}}_k},
	\eeq
	where $\mb{u}_k^{\mr{Up}}$'s be the upper-bound FDPs which can be obtained by solving the following problem,
	\beqn
	\underset{\left\lbrace  \mb{u}_k\right\rbrace }{\max} \;\; \tilde{\eta}(\mb{U})=\dfrac{R^{\mr{FDP}}_{k}(\mb{U})}{\sum_{\forall k} \mb{u}_k^H\mb{u}_k} \;\; \text{ s. t. } \sum_{\forall k} \mb{u}_{k}^{H} \mb{E}_t \mb{u}_{k} \leq P^{\mr{max}}_t \;\;\; \forall (t).  \label{SEE-max-prb_upper2}
	\eeqn
	in which $\mb{U}$ is denoted as the matrix generated by all FDP vectors $\mb{u}_k$'s $\in \mathbb{C}^{N_{\mr{T}} \times 1}$, $\mb{E}_t=diag(0_{1 \times (t-1)},1$ $,0_{1 \times (N_{\mr{T}}-t)}) $, and
	\beq \label{Rate_U}
	R^{\mr{FDP}}_{k}(\mb{U})=\log \left(1 + \frac{ \big|\mb{h}^{H}_{k} \mb{u}_{k} \big|^2}{\sum_{j \neq k} \big|\mb{h}^{H}_{k} \mb{u}_{j} \big|^2 + \sigma^2 }\right).
	\eeq
\end{proposition}
\begin{IEEEproof}
The proof is given in Appendix~\ref{prf_P2_prp1}.
\end{IEEEproof}

\subsubsection{Proposed Framework Solving Problem \eqref{SEE-max-prb_upper2}}
We are now ready to determine the upper-bound of $\eta(\mb{W},\mb{A})$ by proposing a framework solving problem \eqref{SEE-max-prb_upper2} based on EEHP-A algorithm given in \cite{Zi_Jsac_16}.
Denoting the upper-bound of SEE $\tilde{\eta}(\mb{U})$ and the rate in \eqref{Rate_U} as the functions of $\mb{u}_k$, i.e., $\tilde{\eta}(U)$ and $R^{\mr{FDP}}_k(\mb{U})$. The gradient of $\tilde{\eta}(\mb{U})$ with respect to $\mb{u}_k$ is derived by
\begin{subequations} \label{U_update}
\begin{eqnarray}
\dfrac{\partial \tilde{\eta}(\mb{U})}{\partial \mb{u}_k} & = & \dfrac{2}{\tilde{P}_{\mr{T}}^2} \Big(\Upsilon_k - \Phi_k \Big) \mb{w}_k, \\
\text{with } \Upsilon_k &=&\frac{P_{\mr{T}}\mb{h}_k\mb{h}^{H}_k}{\sum_{\forall j}\mb{h}^{H}_k \mb{u}_{j} \mb{u}_{j}^H \mb{h}_k + \sigma^2}, \\
\Phi_k &=& \frac{\sum_{\forall j} R^{\mr{FDP}}_j}{\ln 2}\mb{I} + P_{\mr{T}}\sum_{j \neq k}\frac{p_{k} \mb{h}_k\mb{h}^{H}_k}{m_k(m_k-p_{k})},\\
P_{\mr{T}}&=&\sum_{\forall k} \mb{u}_k^H\mb{u}_k,
\end{eqnarray}
\end{subequations}
where $p_{k}= \mb{h}^{H}_k \mb{u}_{k} \mb{u}_{k}^H \mb{h}_k$ and $m_k=\sum_{\forall j}\mb{h}^{H}_k \mb{u}_{j} \mb{u}_{j}^H \mb{h}_k + \sigma^2$. Then, the local optimization solution for $\mb{u}_k$, can be defined by applying the zero-gradient condition ${\partial \tilde{\eta}(\mb{U})}/{\partial \mb{u}_k}=0$ as $\mb{u}_k = \Phi_k^{-1} \Upsilon_k \mb{u}_k$. 
To obtain the optimal FDP vectors $\mb{u}_k$'s, an iterative algorithm is developed as in Algorithm~\ref{P2_alg:3} which is  similar to the EEHP-A algorithm in \cite{Zi_Jsac_16}. Let $\mb{U}^{\mr{Up}}$ be the optimal solution of problem \eqref{SEE-max-prb_upper2}. Then, for any feasible solution $(\mb{W}^{\prime},\mb{A}^{\prime})$, $\eta(\mb{W}^{\prime},\mb{A}^{\prime})$ is upper-bounded by
$ \bar{\eta}(\mb{U}^{\mr{Up}},\mb{B}^{\prime})$,
where $\mb{B}^{\prime}$ is defined based on $\mb{A}^{\prime}$. 
\begin{algorithm}[!t]
\caption{\textsc{Iterative Weighted Rate and Power Maximization Hybrid Precoding}}
\label{P2_alg:3}
\begin{algorithmic}[1]
\STATE Initialize by choosing any $\mb{U}^{(0)}$ satisfying the power constraint, selecting a tolerance for converging $\tau^{\mr{up}}$, and setting $l=0$.
\REPEAT 
\STATE Update $\Upsilon_k$'s and $\Phi_k$'s as in \eqref{U_update} based on $\mb{U}^{(l)}$.
\STATE Determine the step size $\mu_{k}^{(l+1)}$'s by solving the following problem. 
\begin{subequations} \label{W_update}
\begin{eqnarray}
&& \hspace{-1.7cm} \mu_k^{(l+1)} = \underset{\mu_k \in [0,1]}{\arg \max} \bar{\eta} \Big\{ \big[\mb{I} + \mu_k( \Upsilon_k^{(l)}/\Phi_k^{(l)}-\mb{I}) \big]\mb{u}_k^{l} \Big\} \\
&& \hspace{-1.7cm}  \text{s.t.  } \sum_{\forall k} \Vert \big[\mb{I} + \mu_k(\Upsilon_k^{(l)}/\Phi_k^{(l)}-\mb{I}) \big] \mb{E}_t \mb{u}_k^{l} \Vert^2  \leq P_t^{\mr{max}}, \forall t.
\end{eqnarray}
\end{subequations}
\STATE Update $\mb{u}_k^{(l+1)}=\big[\mb{I} + \mu_k^{(l+1)}( \Upsilon_k^{(l)}/\Phi_k^{(l)}-\mb{I}) \big]\mb{u}_k^{l}$, $\forall l$.
\UNTIL Convergence, i.e., $\vert \tilde{\eta}( \mb{U}^{(l)}) - \tilde{\eta}( \mb{U}^{(l-1)}) \vert \leq \tau^{\mr{up}} $.
\end{algorithmic}
\end{algorithm} }

\subsection{Heuristic Energy Efficiency Maximization Method}
In this section, we first re-construct a near-optimal performance HP based on $\mb{U}^{\mr{Up}}$ \cite{Ayach2014} for a given $\mb{B}$. Then, we develop an heuristic algorithm for solving problem \eqref{SEE-max-prb}.
\subsubsection{Hybrid Precoding Design for Given $\mathbf{B}$} \label{solveAW_fixedB}
We aim to reconstruct the HP where $\mb{w}_{k}$'s and $\mb{A}$ can be defined via MMSE approximation as follows.
\beq \label{HP_givenB}
\underset{\mb{W},\mb{A}}{\min} \sum_{\forall k} \Vert \mb{u}^{\mr{Up}}_{k} - \mb{A} \mb{w}_{k} \Vert_2^2 \;\; \text{s. t. \eqref{cnt2} and \eqref{a_b_related}}.
\eeq
For given $\mb{A}$, $\mb{w}_k$'s can be determined based on the well-known least squares method as
\beq
\mb{w}_{k} =  \beta_k \mb{A}^{\dagger} \mb{u}^{\mr{Up}}_{k},
\eeq
where $\beta_k$'s are the factors satisfying the power constraint \eqref{cnt2}.
While fixing $\mb{w}_{k,s}$'s, $\mb{A}$ can be obtained by solving the following problem:
\begin{eqnarray} \label{frobenius_prob_A}
\underset{\mb{A}}{\min}   \sum \limits_{\forall k} \big\|\mb{u}^{\mr{Up}}_{k} - \mb{A} \mb{w}_{k}\big\|^2_2 \quad
\text{ s. t. \eqref{a_b_related}}.
\end{eqnarray}
This problem is classified as a unit-modulus least square type, which is non-convex and NP-hard. This problem can be efficiently solved by employing the ``Projected Gradient Descent Method'' proposed in \cite{VuHa_TWC18}.
\subsubsection{A Heuristic Algorithmic Approach}
Here, we develop an heuristic algorithm to solve the SEEM problem.
First, we define the upper-bound FDP as in Algorithm~\ref{P2_alg:3}.
Then, we start with $\mb{B}=\mathbf{1}_{N_{\mr{T}} \times N_{\mr{RF}}}$.
In each iteration, we define $(\mb{W},\mb{A})$ based on $\mb{B}$ and $\mb{U}^{\mr{Up}}$ before deactivating the active connection without which one can increase the SEE most.
The process stops if no activated connection can be turned off to increase the SEE.

\begin{algorithm}[!t]
\caption{\textsc{Heuristic Energy-Efficiency Maximization Hybrid Precoding Algorithm}}
\label{P2_alg:4}
\begin{algorithmic}[1]
\STATE Initialize: Define $\mb{U}^{\mr{Up}}$ by employing Algorithm~\ref{P2_alg:3}. Set $\mb{B}=\mathbf{1}_{N_{\mr{T}} \times N_{\mr{RF}}}$ and $\mathcal{L}=\lbrace(t,n)\vert b_t^n =1\rbrace$.
\REPEAT 
\STATE Solve $\mb{W}$ and $\mb{A}$ based on $\mb{U}^{\mr{Up}}$ and $\mb{B}$ as discussed in Section~\ref{solveAW_fixedB}.
\STATE For every $(t,n) \in \mathcal{L}$, define $\mb{A}^{(t,n)} = \mb{A}$ then set $(\mb{A}^{(t,n)})_{t,n}=0$.
\STATE Define $(t^{\prime},n^{\prime})=\arg \max_{(t,n) \in \mathcal{L}} \eta(\mb{W},\mb{A}^{(t,n)})$.
\IF {$\eta(\mb{W},\mb{A}^{(t^{\prime},n^{\prime})}) > \eta(\mb{W},\mb{A})$}
\STATE Set $\mathcal{L}=\mathcal{L}/(t^{\prime},n^{\prime})$ and update $(\mb{B})_{t^{\prime},n^{\prime}}=0$. 
\ENDIF
\UNTIL $\eta(\mb{W},\mb{A}^{(t^{\prime},n^{\prime})}) \leq \eta(\mb{W},\mb{A})$.
\end{algorithmic}
\end{algorithm}

\section{Simulation Results} \label{sec5}
\subsection{mmWave Channel Model}
The mmWave channel is generally not rich in scattering because mmWave signals do not reflect well in the surrounding environment \cite{Rappaport_book}. Hence, there are only few dominant paths in mmWave transmission channel. Similar to \cite{Yu_TSP_16}, the Saleh-Valenzuela geometric channel model is adopted for the numerical evaluation in this paper as
\beq
\mb{h}_{k}  = \sqrt{\frac{N_{\mr{T}}}{P_{\mr{L}} N_{\mr{C}} N_{\mr{L}}}}  \sum_{c=1}^{N_{\mr{C}}} \sum_{\ell=1}^{N_{\mr{L}}} \alpha_{c,\ell} a_{\mr{r}}(\phi_{c,\ell}^{\mr{r}},\theta_{c,\ell}^{\mr{r}}) \mb{a}_{\mr{t}}(\phi_{c,\ell}^{\mr{t}},\theta_{c,\ell}^{\mr{t}}),
\eeq
where $P_{\mr{L}}$ is the path-loss, and $N_{\mr{C}}$ and $N_{\mr{L}}$ are the number of clusters and number of propagation sub-paths in each cluster, respectively. In addition, $\alpha_{c,\ell}$ is the complex gain of the $\ell$-th path of cluster $c$,  and $(\phi^{\mr{r}}_{c,\ell},\theta^{\mr{r}}_{c,\ell})$ and $(\phi^{\mr{t}}_{c,\ell},\theta^{\mr{t}}_{c,\ell})$ are
the (azimuth, elevation) angles of arrival and departure corresponding, respectively. Herein, $a_{\mr{r}}(\phi_{c,\ell}^{\mr{r}},\theta_{c,\ell}^{\mr{r}})$ and $\mb{a}_{\mr{t}}(\phi_{c,\ell}^{\mr{t}},\theta_{c,\ell}^{\mr{t}})$
represent the normalized receive response factor and transmit array response vectors at (azimuth, elevation) angles of $(\phi_{c,\ell}^{\mr{r}},\theta_{c,\ell}^{\mr{r}})$ and $(\phi_{c,\ell}^{\mr{t}},\theta_{c,\ell}^{\mr{t}})$, respectively \cite{Yu_TSP_16,Rappaport_book}. 
Finally, $\alpha_{c,\ell}$ is assumed to be i.i.d. Gaussian distributed and the normalization factor $\sqrt{{N_{\mr{T}}}/{(P_{\mr{L}}N_{\mr{C}}N_{\mr{L}})}}$
is added to get $\mathbb{E}_{\mb{h}_{k}}\big\{\Vert \mb{h}_{k} \Vert_2^2\big\} = N_{\mr{T}}/P_{\mr{L}}$. 

\subsection{Simulation Results}
This section presents the performance of the proposed energy-efficiency HP algorithm which is Algorithm~\ref{P2_alg:1} integrated with Algorithm~\ref{P2_alg:2} (denoted ``Proposed Alg.'' in the figures). For comparison purpose, the performances of other precoding designs are also presented, including: i.) the upper bound corresponding to the FDP achieved by Algorithm~\ref{P2_alg:3} presented in Section \ref{sec:upper}, denoted as ``FDP Upper Bound'', ii.) the heuristic energy-efficiency HP achieved by Algorithm \ref{P2_alg:4}, denoted as ``Heuristic Alg.'', iii.) two algorithms algorithm given in \cite{Zi_Jsac_16} and \cite{He_Tsp_17}, denoted as ``Zi's Alg.'' and ``He's Alg.'', respectively.
An uniform polar array with antenna spacing equal to a half-wavelength is adopted at the base station.
The channel to each user contains three clusters of $10$ paths, i.e., $C=3$, $L = 10$.
All the channel path gains $\alpha_{c,\ell}$'s are assumed to be i.i.d. Gaussian random variables
with variance $\sigma_{\alpha}^2$. 
The azimuth angles are assumed to be uniformly
distributed in $[0; 2\pi]$, and the AoA/AoD elevation angles are uniformly 
distributed in $[-\frac{\pi}{2};\frac{\pi}{2}]$. 
{\color{blue}The noise variance $\sigma^2$ is set at $1.2 \times 10^{-13}$~W due to the noise factor of $4.79$ dB, the power spectral density of $-174$ dBm/Hz, and $1$ MHz bandwidth \cite{TI_2014}.}
In this simulation, we employ the \textit{path loss ABG model} for macro-cellular scenario with $f=60$~GHz in Table I of \cite{Sun_VTC16S}, where distance range is set equal to $100$~m for all users.
{\color{blue}In addition, we set $P_{\mr{BB}} = 300$~mW, $P_{\mr{DAC}} = 200$~mW, $P_{\mr{RFC}} = 43$~mW, $P_{\mr{PS}} = 40$~mW, $P_{\mr{SW}} = 2$~mW, and $P_{\mr{amp}} = 40$~mW, $G_{\mr{amp}} = 20$~dB, $\rho_{\mr{pa}} = 0.3$, $L_{\sf{sw}} = L_{\sf{ps}} = 2$~dB \cite{Rial15,Li_TMTT13,V_Jamali_ICC19}.
For implementing the algorithms, we choose $\varepsilon = 10^{-8}$, $\tau^{\mr{out}}=\tau^{\mr{in}}=\tau^{\mr{up}} = 10^{-4}$.}
Unless indicated otherwise, $P_t^{\mr{max}}$ is set as $50$~mW, both $K$ and $N_{\mr{RF}}$ are set equal to $8$, and $N_{\mr{T}}$ is $64$. 

\begin{figure}[t!]
	\centering
	\includegraphics[width=120mm]{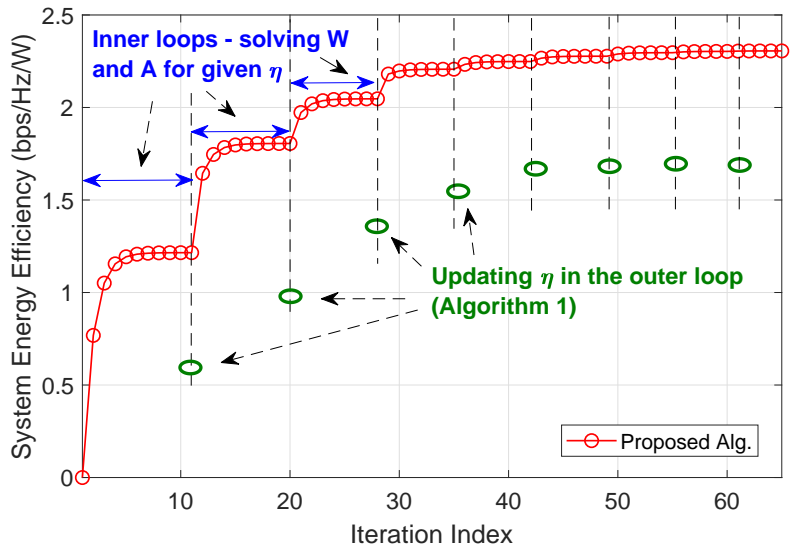}
	\caption{Convergence of the proposed algorithm.}
	\label{Convergence-fig}
\end{figure}

{\color{blue}We illustrate the convergence of our proposed algorithm in Fig.~\ref{Convergence-fig} where the variations of SEE calculated based on the outcomes $(\mb{W},\mb{A})$ of Algorithm~\ref{P2_alg:1} integrating with Algorithm~\ref{P2_alg:2} over the iterations are shown. 
As can be seen, the SEE increases monotonically after each iteration before reaching its the maximum value which is the outcome of Algorithm~\ref{P2_alg:1}.
Additionally, there are many fragments of the iteration sequence within of which the SEE firstly boosts speedily, then grows slower before saturating. This is because of that the iterations in each such fragment are due to the inner loop employing Algorithm~\ref{P2_alg:2} to solve WSRP for given value of $\eta$.
At the end of each fragment, $\eta$ is updated according to Algorithm~\ref{P2_alg:1} before the next inner loop is implemented in the next fragment.
This simulation result hence confirms the convergence of the proposed algorithm proved in Theorem \ref{P2:prp_cvgAlg1} and Proposition \ref{P2_prp_cvgAlg1_Alg2}. }

\begin{figure}[t!]\centering
	\includegraphics[width=120.00mm]{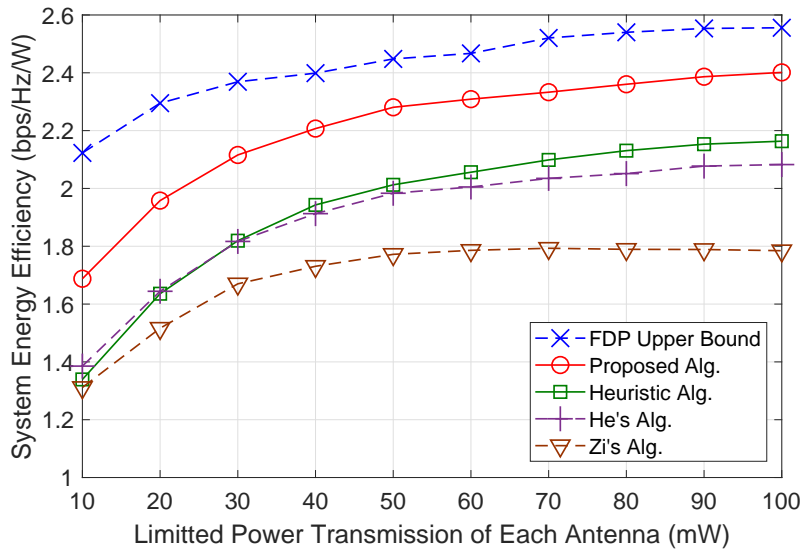}
	\caption{The system energy efficiency \emph{versus} the power budget of each antenna.}
	\label{EEvPt-fig}
\end{figure}

Fig.~\ref{EEvPt-fig} presents the SEEs achieved by the proposed algorithm and other methods versus the transmit power at each antenna, $P_t^{\mr{max}}$. As can be seen, the SEEs achieved by all algorithms increase as $P_t^{\mr{max}}$ increases before saturating at the high regime of $P_t^{\mr{max}}$ ,while the SEE upper bound increases then slightly decreases. This is due to the larger feasible set that can attain better solutions. However, the FDP is designed to maximize the data rate to the transmit power ratio, not the data rate to the total power consumption ratio. Hence, when the increase of $P_t^{\mr{max}}$ may result in more active phase shifters which then reduces the SEE. In addition, the proposed algorithm achieves much higher achievable SEEs at all values of $P_t^{\mr{max}}$ than the heuristic algorithm and the two methods given in \cite{Zi_Jsac_16,He_Tsp_17} do since our design takes care of all the total consumption power while the works in \cite{Zi_Jsac_16,He_Tsp_17} do not.
Interestingly, the method given in \cite{Zi_Jsac_16} obtains the smallest SEE in comparison to others while the heuristic algorithm outperforms He's algorithm in \cite{He_Tsp_17} when value of $P_t^{\mr{max}}$ increases. 
	
\begin{figure}[t!]\centering
	\includegraphics[width=120.00mm]{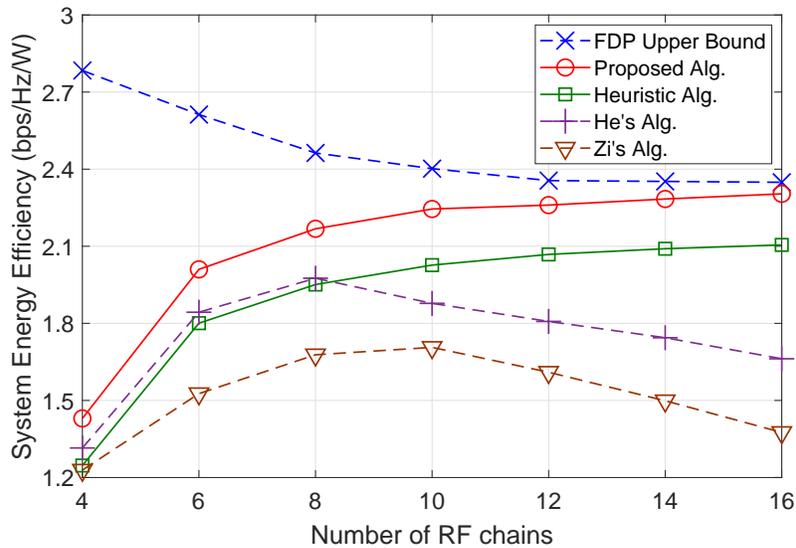}
	\caption{The system energy efficiency \emph{versus} the number of RF chains $N_{\mr{RF}}$.}
	\label{EEvRF-fig}
\end{figure}

Fig.~\ref{EEvRF-fig} illustrates the SEE achieved by all schemes versus the number of RF chains, $N_{\mr{RF}}$.
As can be observed, the SEE upper bound (in Section \ref{sec:upper}) decreases as $N_{\mr{RF}}$ increases since the achievable rate is unchanged while the system consumes more power. When $N_{\mr{RF}}$ becomes larger, the proposed algorithms can achieve better SEE. In contrast, the SEE obtained by Zi's and He's methods first increases with $N_{\mr{RF}}$ then decreases. 
Interestingly, the proposed algorithm significantly outperforms the other HP designs and achieves a SEE near to the upper bound at the high regime of $N_{\mr{RF}}$, which again confirms the superior performance of the proposed design.

\begin{figure}[t!]\centering
	\includegraphics[width=120.00mm]{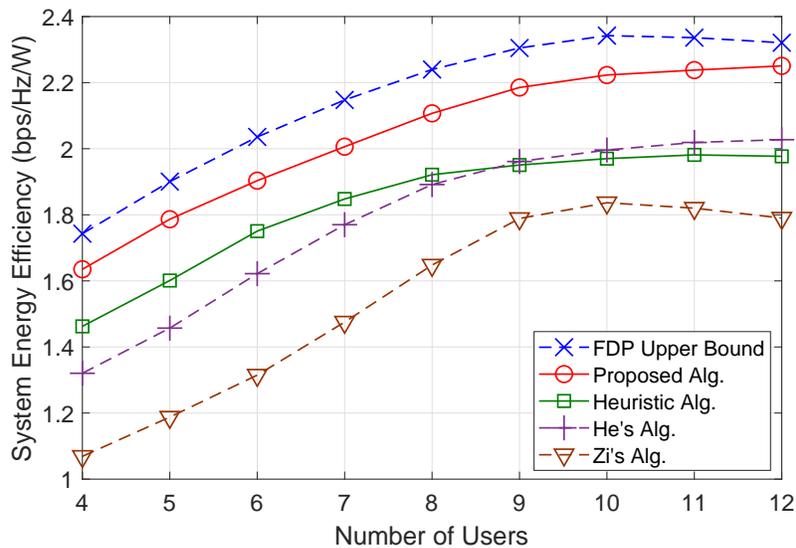}
	\caption{The system energy efficiency \emph{versus} the number of users $K$.}
	\label{EEvK-fig}
\end{figure}

In Fig.~\ref{EEvK-fig}, the SEE achieved by all schemes are displayed versus the number of users $K$.  The SEE achieved by all schemes increases and then decreases with the number of users. This observation is arguably due to the increased  power consumption for baseband signal processing to support more users.  Furthermore, the increase rate in the required power is much faster than the rate in the sum-rate improvements. Again, our proposed algorithm can attain better SEE than other HP design methods.
  
\begin{figure}[t!]\centering
	\includegraphics[width=120.00mm]{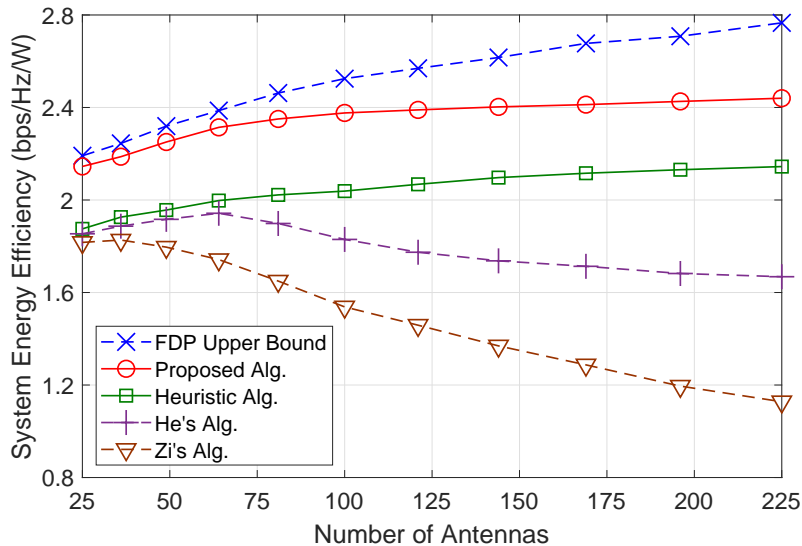}
	\caption{The system energy efficiency \emph{versus} the number of antennas $N_{\mr{T}}$.}
	\label{EEvNT-fig}
\end{figure}

\begin{figure}[t!]\centering
	\includegraphics[width=120.00mm]{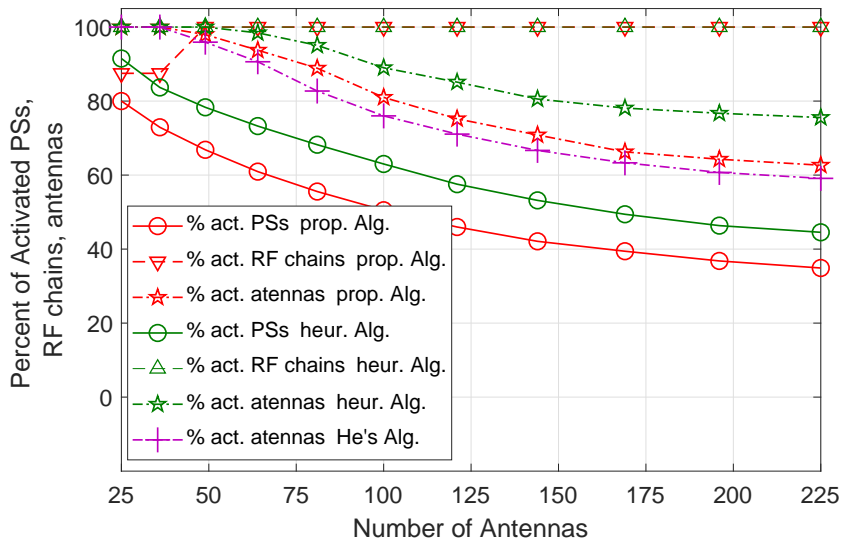}
	\caption{The percentage of activated phase shifters, RF chains, and antennas \emph{versus} the number of antennas $N_{\mr{T}}$.}
	\label{ActivNT-fig}
\end{figure}

\begin{figure}[t!]\centering
	\includegraphics[width=120.00mm]{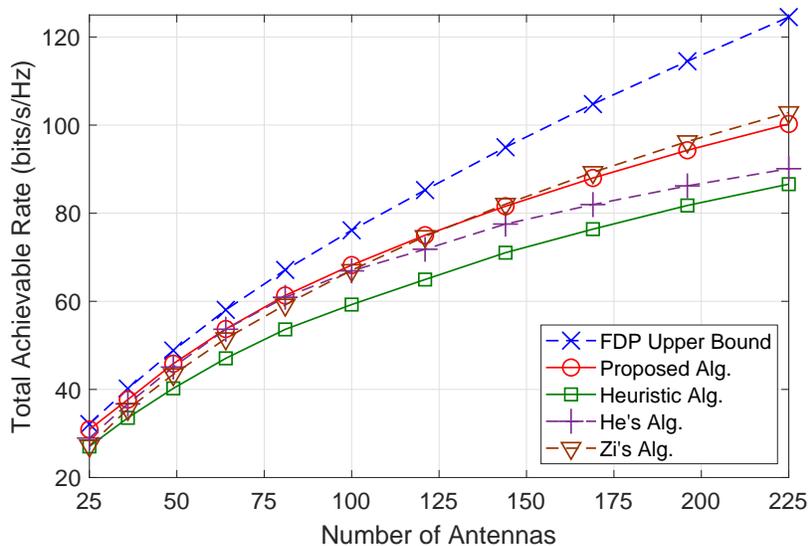}
	\caption{Total achievable rate \emph{versus} the number of antennas $N_{\mr{T}}$.}
	\label{RatevNT-fig}
\end{figure}

Next, the impact of the number of antennas $N_{\mr{T}}$ is presented in Figs.~\ref{EEvNT-fig}, \ref{ActivNT-fig}, and \ref{RatevNT-fig}. Fig. \ref{EEvNT-fig} illustrates the SEEs achieved by different schemes  while varying $N_{\mr{T}}$. While the upper bound keeps increasing with $N_{\mr{T}}$, the proposed algorithm increases then saturates at large $N_{\mr{T}}$. On the other hand, the SEE achieved by the two methods given in \cite{Zi_Jsac_16} and \cite{He_Tsp_17} increases then deceases quickly. In this simulation, our proposed method again outperforms all other designs except the upper bound by the FDP.

In Fig. \ref{ActivNT-fig}, we present the percentage of phase shifters, RF chains, and antennas which are activated with the proposed algorithm and the heuristic algorithm in scenarios with different number of antennas. The percentage of utilized antennas with the method given in \cite{He_Tsp_17} is also illustrated for comparison. Interestingly, the percentage of activated phase shifters and antennas decreases as $N_{\mr{T}}$ increases while the percentage of activated RF chains increases. As can be observed, the proposed algorithm turns off more phase shifters, RF chains, and antennas than the heuristic one. In addition, He's algorithm utilizes fewer antennas than others since this HP design focuses on reducing the number of antennas to maximize the SEE.

Fig.~\ref{RatevNT-fig} presents the achievable rate achieved by all schemes versus the number of antennas $N_{\mr{T}}$. As expected, the system achievable rate achieved by all schemes increases with $N_{\mr{T}}$.
The FDP enhances well the ``degree-of-freedom'' to achieve the highest data rate.
Zi's algorithm place itself in the second place since this scheme utilizes all the hardware components to serve all users. 
He's algorithm outperforms our proposed algorithm in the low regime of $N_{\mr{T}}$ while ours is superior to this method in the high regime.
In addition, the proposed algorithm is better than the heuristic one in all scenarios with different number of antennas.

\section{Conclusion} \label{ccls}
This paper has considered the dynamic fully-connected AP structure for the HP mmWave multi-user systems where each connection between one RF chain and one antenna can be activated or deactivated for power saving. This dynamic fully-connected AP structure has been represented as a sparsity-modulus constraint for the AP matrix design. In addition, the total power consumption of this system is formulated generally as the sparsity form of AP matrix. Then, a new compressed sensing-based energy-efficiency HP algorithm has been proposed to maximize the non-convex SEE.
Numerical results have confirmed the superior performances of the proposed energy-efficiency HP design over heuristic algorithms and benchmark algorithms in literature.

\appendices
\section{Proof of Theorem~\ref{thr_eta_star}}
\label{prf_thr_eta_star}
{\color{blue}Theorem~\ref{thr_eta_star} can be proved by employing an approach similar to \cite{Dinkelbach67} as follows.
\begin{itemize}
\item[i)] Consider any $\eta_1$ and $\eta_2$ where $0< \eta_1 < \eta_2$. Let $\mb{x}_1$ and $\mb{x}_2$ be the optimal solution of problem~\eqref{sub-x-prob} corresponding to $\eta_1$ and $\eta_2$, respectively. Then, we have:
\beqn
\chi(\eta_2) & = & R(\mb{x}_2)- \eta_2 P(\mb{x}_2) \nonumber \\
& < & R(\mb{x}_2)- \eta_1 P(\mb{x}_2) \nonumber \\
& \leq & R(\mb{x}_1)- \eta_1 P(\mb{x}_1) = \chi(\eta_1)
\eeqn
Therefore, $\chi(\eta_2)<\chi(\eta_1)$ for any $\eta_1$ and $\eta_2$ where $0< \eta_1 < \eta_2$, which means $\chi(\eta)$ is a strictly monotonic decreasing function.
\item[ii)] Assume that $\chi(\eta) > 0$. Then, there exists $\hat{\mb{x}}$ such that $R(\hat{\mb{x}})- \eta P(\hat{\mb{x}}) > 0$. Hence, we have $\eta < \frac{R(\hat{\mb{x}}}{P(\hat{\mb{x}})}$, which implies that 
\beq
\eta < \max_{\mb{x} \in \mathcal{S}} \frac{R(\mb{x})}{P(\mb{x})} = \eta^{*}.
\eeq
Conversely, if $\eta < \eta^{*}$, then, we have $\eta < \max_{\mb{x} \in \mathcal{S}} \frac{R(\mb{x})}{P(\mb{x})}$. Hence, there exists $(\hat{\mb{x}}')$ such that $\eta < \frac{R(\hat{\mb{x}}')}{P(\hat{\mb{x}}')}$, which implies that 
\beq
\chi(\eta) = \max_{\mb{x} \in \mathcal{S}} R(\mb{x})- \eta P(\mb{x}) >0.
\eeq
\item[iii)] Assume that $\chi(\eta) < 0$. Then, we have $R(\mb{x})- \eta P(\mb{x}) < 0$ $\forall \mb{x} \in \mathcal{S}$. Hence, $\eta > \frac{R(\mb{x})}{P(\mb{x})}$ $\forall \mb{x} \in \mathcal{S}$, which yields
\beq
\eta > \max_{\mb{x} \in \mathcal{S}}\frac{R(\mb{x})}{P(\mb{x})} = \eta^{\star}.
\eeq
Conversely, if $\eta > \eta^{\star}$, then, we have $\eta > \max_{\mb{x} \in \mathcal{S}}\frac{R(\mb{x})}{P(\mb{x})}$. Hence, $\eta > \frac{R(\mb{x})}{P(\mb{x})}$ $\forall \mb{x} \in \mathcal{S}$, which implies 
\beq
\chi(\eta) = \max_{\mb{x} \in \mathcal{S}} R(\mb{x})- \eta P(\mb{x}) <0.
\eeq
\item[iv)] Due to the above results, one has $\chi(\eta^{\star}) = 0$. Then, if $\mb{x}^{*}$ is an optimal solution of the problem $(\mathcal{P}_I)$, this implies $R(\mb{x}^{*})- \eta^{*} P(\mb{x}^{*}) = 0$. Thus, $\mb{x}^{*}$ is also an optimal solution of $(\mathcal{P}_{II}^{\eta^{\star}})$. 

Conversely, if $\mb{x}^{\prime}$ is an optimal solution of $(\mathcal{P}_{II}^{\eta^{*}})$, one has that $\chi(\eta^{*}) = R(\mb{x}^{\prime})- \eta^{*} P(\mb{x}^{\prime})$ must equal to $0$ due to the observation (ii) and (iii). 
Hence, we have $\frac{ R(\mb{x}^{\prime})}{P(\mb{x}^{\prime})} = \eta^{*}$, which yields that $\mb{x}^{\prime}$ is an optimal solution of problem $(\mathcal{P}_I)$. Thus, $(\mathcal{P}_I)$ and $(\mathcal{P}_{II}^{\eta^{\star}})$ have the same set of optimal solutions.
\end{itemize}}

\section{Proof of Theorem~\ref{P2:prp_cvgAlg1}}
\label{prf_P2_prp_cvgAlg1}
We will prove that $\eta^{(m)}$ monotonically increases after each iteration.
It is easy to see that if $\bar{\chi}(\eta^{(m)},\mb{W}^{(m)},\mb{A}^{(m)}) \geq \bar{\chi}(\eta^{(m)},\mb{W}^{(m-1)},\mb{A}^{(m-1)})$, one has
\beqn
 R(\mb{W}^{(m)},\mb{A}^{(m)}) - \eta^{(m)} P_{\mr{tot}}(\mb{W}^{(m)},\mb{A}^{(m)}) \geq R(\mb{W}^{(m-1)},\mb{A}^{(m-1)}) - \eta^{(m)} P_{\mr{tot}}(\mb{W}^{(m-1)},\mb{A}^{(m-1)}) = 0,  
\eeqn
since $\eta^{(m)}={R(\mb{W}^{(m-1)},\mb{A}^{(m-1)})}/{P_{\mr{tot}}(\mb{W}^{(m-1)},\mb{A}^{(m-1)})}$. Hence, the process updating $\eta^{(m+1)}$ given in Step 4 of Algorithm~\ref{P2_alg:1} can imply that 
\beq
\eta^{(m+1)}=\frac{R(\mb{W}^{(m)},\mb{A}^{(m)})}{P_{\mr{tot}}(\mb{W}^{(m)},\mb{A}^{(m)})} \geq \eta^{(m)}.
\eeq
The result implies the monotonic increase of $\eta$ after each iteration.
Furthermore, the value of energy efficiency $\eta$ cannot be infinity.
Hence, the Algorithm~\ref{P2_alg:1} can converge after a finite number of iterations.

\section{Proof of Theorem~\ref{P2_thr2} }
\label{prf_P2_thr2}
According to the receive coefficient $\delta_k$, the estimate of $s_k$ at the user $k$ can be given by $\hat{s}_k=\delta_{k} y_{k}$. Hence, the MMSE  receive coefficient at user $k$ is given by

\begin{eqnarray} \label{receive}
\delta_{k}^{\mr{MMSE}} = \arg\min_{\delta_{k}} \mathbb{E}\big\{\left|s_{k} - \delta_{k} y_{k}\right|^2\big\}  = \bigg(\sum_{\forall j} \vert\mb{h}_{k}^{H} \mb{A} \mb{w}_j\vert^2 +  \sigma_k^2 \bigg)^{-1}\mb{w}_k^H \mb{A}^H \mb{h}_{k}. 
\end{eqnarray}
Then, the MSE for user $k$ corresponding to the MMSE-receive filter can be written as
$ e_k = \mathbb{E}\big\{\left|s_{k} - \delta_{k}^{\mr{MMSE}} y_{k}\right|^2\big\} = \big(1 + \mr{SINR}_k \big)^{-1}$.
Hence, $R_k(\mb{A},\mb{W})$ can be expressed as a function of
$e_k$ as $R_k(\mb{A},\mb{W})=\log(e_k^{-1})$. Furthermore, thanks to the first-order Taylor approximation for the $\log$-function, one can yield
\beqn
R_k(\mb{A},\mb{W})&&=\log(e_k^{-1})=-\log(e_k) \geq - \big[\log(\omega^{-1}_k)+\omega^{-1}_k(e_k -\omega_k)\big]. 
\eeqn
In addition, it is easy to see that the optimum value of $\delta_k$ and $\omega_k$ can be expressed as
\beqn
\delta_k^{\star}  =  \delta_{k}^{\mr{MMSE}} \text{ and } \omega_k^{\star} = e_k^{-1} \label{omega}.
\eeqn
At that point one has $g(\mb{A},\mb{W},\delta_{k},\omega_{k})=l(\mb{A},\mb{W})$. The proof thus follows.

\section{Proof of Theorem~\ref{prp_cvgAlg2}} \label{prf_P2_prp_cvgAlg2}
Denote the $a_t^{n(t)}$ as the solution of $a_t^n$ at the $\ell$th iteration in Algorithm~\ref{P2_alg:2}. The majorization of  the $\log$ function given in \eqref{log_maj} in conjunction with updating $\delta_{k}^{[\ell]}$'s in Step 4 (as in \eqref{receive2}), determining $\omega_{k}^{[\ell]}$'s in Step 5 (as in \eqref{omega2}), optimizing $\mb{w}_{k}^{[\ell]}$'s in Step 6 (by solving problem (\ref{QCQP-probW})), and solving problem (\ref{QCQP-probAt})'s to obtain $\mb{a}_{t}^{[\ell]}$ in Step 7--9 ensure the decrement of the objective function in problem \eqref{WMMSE-prob} at the $\ell$th iteration. Then, we have
\beqn
g^{[\ell+1]}&&=\Sigma^{[\ell+1]} + \sum \limits_{\forall X^{[\ell+1]}} \eta_{X}\log(X^{[\ell+1]}+\varepsilon) \nonumber \\
&&\leq \Sigma^{[\ell+1]} + \sum \limits_{\forall X} \eta_{X} \bigg[ \log(X^{[\ell]}+\varepsilon) + \frac{X^{[\ell+1]}-X^{[\ell]}}{X^{[\ell]}+\varepsilon} \bigg] \nonumber\\
&&\leq \Sigma^{[\ell]} + \sum \limits_{\forall X} \eta_{X} \bigg[ \log(X^{[\ell]}+\varepsilon) + \frac{X^{[\ell]}-X^{[\ell]}}{X^{[\ell]}+\varepsilon} \bigg] \nonumber\\
&&\leq \Sigma^{[\ell]} + \sum \limits_{\forall X} \eta_{X} \log(X^{[\ell]}+\varepsilon) = g^{[\ell]},
\eeqn
where $g^{[\ell]}$ stands for the value of the objective function in problem~\eqref{WMMSE-prob2} at the $\ell$th iteration.
Therefore, the convergence of Algorithm~\ref{P2_alg:2} is attained thanks to the monotonic decrease of the objective function in problem \eqref{WMMSE-prob2} after each iteration.

\section{Proof of Proposition~\ref{P2_prp_cvgAlg1_Alg2}}
\label{prf_P2_prp_cvgAlg1_Alg2}
In the $m$-th iteration of the outer loop, Algorithm~\ref{P2_alg:2} is called for solving the WRSP problem for a given $\eta^{(m)}$. The starting point of the inner loop is chosen as $(\mb{W}^{(m-1)},\mb{A}^{(m-1)})$, which is the outcome of the $(m-1)$-th iteration of the outer loop. Thanks to Theorem~\ref{prp_cvgAlg2}, the objective function of problem~\eqref{WMMSE-prob2} monotonically decreases from the initial point to a convergence point.
This convergence point is the set as the outcome of the $m$-th iteration of the outer loop. Hence, we have
\beq
g(\mb{W}^{(m)},\mb{A}^{(m)},\delta_k^{(m)},\omega_k^{(m)}) \leq g(\mb{W}^{(m-1)},\mb{A}^{(m-1)},\delta_k^{(m-1)},\omega_k^{(m-1)}).
\eeq
This result combined with Theorem~\ref{P2_thr2} yields
\beq
\bar{\chi}(\eta^{(m)},\mb{W}^{(m)},\mb{A}^{(m)}) \geq \bar{\chi}(\eta^{(m)},\mb{W}^{(m-1)},\mb{A}^{(m-1)}). \nonumber
\eeq
Proposition~\ref{P2_prp_cvgAlg1_Alg2} thus follows.

{\color{blue}\section{Proof of Proposition~\ref{P2_prp1}}
\label{prf_P2_prp1}

Denote $(\mb{A}^{\prime},\mb{W}^{\prime})$ as an arbitrary feabible solution of problem \eqref{SEE-max-prb}.
Let $\mb{u}^{\prime}_k = \mb{A}^{\prime}\mb{w}_k^{\prime}$. Then, one has 
\beq
\sum_{\forall k} \mb{u}_{k}^{\prime H} \mb{E}_t \mb{u}^{\prime}_{k} = \mb{w}^{\prime H}_{k}\mb{a}^{\prime}_t\mb{a}_t^{\prime H}\mb{w}^{\prime}_{k} \leq P^{\mr{max}}_t \;\;\; \forall (t),
\eeq
which yields $\mb{u}^{\prime}_k$'s is a feasible solution of problem \eqref{SEE-max-prb_upper2}.
It means
\beqn
\tilde{\eta}(\mb{U}^{\prime}) & = & \dfrac{\sum_{\forall k} \log \left(1 + \frac{ \big|\mb{h}^{H}_{k} \mb{A}^{\prime} \mb{w}^{\prime}_{k} \big|^2}{\sum_{j \neq k} \big|\mb{h}^{H}_{k} \mb{A}^{\prime} \mb{w}^{\prime}_{j} \big|^2 + \sigma^2 }\right)}{\sum_{\forall k} \mb{w}_k^{\prime H} \mb{A}^{\prime H} \mb{A}^{\prime} \mb{w}^{\prime}_k} \nonumber \\
& \leq & \dfrac{\sum_{\forall k} \log \left(1 + \frac{ \big|\mb{h}^{H}_{k} \mb{u}^{\mr{Up}}_{k} \big|^2}{\sum_{j \neq k} \big|\mb{h}^{H}_{k} \mb{u}^{\mr{Up}}_{j} \big|^2 + \sigma^2 }\right)}{\sum_{\forall k} \mb{u}_k^{\mr{Up} H}\mb{u}^{\mr{Up}}_k} = \tilde{\eta}(\mb{U}^\mr{Up}), \quad \forall (\mb{W}^{\prime},\mb{A}^{\prime}) \in \Theta, \label{upbound1}
\eeqn
where $\mb{U}^{\prime}$ which is the matrix generated by all vectors $\mb{u}^{\prime}_k$'s.
Thanks to \eqref{upbound1}, one can conclude that
\beqn
\eta(\mb{W},\mb{A}) & =& \dfrac{\sum_{\forall k} \log(1 + \mr{SINR}_{k})}{P_{\mr{cons}}+\bar{P}_{\mr{spr}}(\mb{B})+(1+1/\rho_{\mr{pa}})\sum_{\forall k} \mb{w}_k^H\mb{A}^H\mb{A}\mb{w}_k} \nonumber \\
& \leq & \dfrac{\sum_{\forall k} R^{\mr{FDP}}_{k}(\mb{U}^{\mr{Up}})}{P_{\mr{cons}}+\bar{P}_{\mr{spr}}(\mb{B})+ (1+1/\rho_{\mr{pa}}) \sum_{\forall k} \mb{u}_k^{\mr{Up} H}\mb{u}^{\mr{Up}}_k} = \bar{\eta}(\mb{U}^\mr{Up},\mb{B}), \quad \forall (\mb{W},\mb{A}) \in \Theta, \label{upbound2}
\eeqn
which finised the proof of Proposition~\ref{P2_prp1}.}

\bibliographystyle{IEEE}
\bibliography{am_ger_eng,rubi_eng}

%
%
%
%
%
%
%
%
\end{document}